\documentclass{article}

\usepackage{graphicx}
\usepackage{balance}
\usepackage[T1]{fontenc}
\usepackage{url}
\usepackage[draft=true]{minted}
\usepackage{amssymb}
\usepackage{tablefootnote}
\usepackage{booktabs}
\usepackage{color, colortbl}

\newcommand{\sccomment}[1] { \emph{ << #1 >> } }
\newcommand{\stackcacheentry}[3]{
\makebox[3.5mm][r]{\scriptsize #1}
\makebox[3.5mm]{\scriptsize #2}
\makebox[2.5mm][l]{\tiny #3}
}
\newcommand{\myfiguresizeperformance}[0] { 6.7cm }
\newcommand{\myfiguresizecodesize}[0] { 6.7cm }
\newcommand{\myfiguresizexxtea}[0] { 6.7cm }
\newcommand{\xt}[0] { \tiny }
\newcommand{\xxt}[0] { \makebox[2mm]{} }
\newcommand{\xxxt}[0] { \makebox[5mm]{} \tiny }
\definecolor{yellow}{rgb}{1,1,0.8}
\newcommand{\tblhighlight}[0] {\cellcolor{yellow}}

\begin{document}

\title{Improved Ahead-of-Time Compilation of Stack-Based JVM~Bytecode on Resource-Constrained Devices}  
\author{Niels Reijers, Chi-Sheng Shih}
\date{NTU-IoX Research Center \\
Department of Computer Science and Information Engineering \\
National Taiwan University}

\begin{abstract}
Many virtual machines exist for sensor nodes with only a few KB RAM and tens to a few hundred KB flash memory. They pack an impressive set of features, but suffer from a slowdown of one to two orders of magnitude compared to optimised native code, reducing throughput and increasing power consumption.

Compiling bytecode to native code to improve performance has been studied extensively for larger devices, but the restricted resources on sensor nodes mean most modern techniques cannot be applied. Simply replacing bytecode instructions with predefined sequences of native instructions is known to improve performance, but produces code several times larger than the optimised C equivalent, limiting the size of programmes that can fit onto a device.

This paper identifies the major sources of overhead resulting from this basic approach, and presents optimisations to remove most of the remaining performance overhead, and over half the size overhead, reducing them to 69\% and 91\% respectively. While this increases the size of the VM, the break-even point at which this fixed cost is compensated for is well within the range of memory available on a sensor device, allowing us to both improve performance and load more code on a device.
\end{abstract}

\maketitle

\section{Introduction}
Internet-of-Things devices come in a wide range, with vastly different performance characteristics, cost, and power requirements. On one end of the spectrum are devices like the Intel Edison and Raspberry Pi: powerful enough to run Linux, but relatively expensive and power hungry. On the other end are CPUs like the Atmel Atmega or TI MSP430, commonly used in sensor nodes: much less powerful, but also much cheaper and low power enough to potentially last for months or years on a single battery. For the first class normal operating systems, languages, and compilers can be used, but in this paper, we focus specifically on the latter class for which no such clear standards exist. Our experiments were all performed on an ATmega128: a 16MHz 8-bit processor, with 4KB of RAM and 128KB of flash programme memory, but the approach should yield similar results on other CPUs in this category.

There are several advantages to using VMs. One is ease of programming. Many VMs allow the developer to write programmes at a higher level of abstraction than the bare-metal C programming that is still common for these devices. Second, a VM can offer a safe execution environment, preventing buggy or malicious code from disabling the device. A third advantage is platform independence. While early wireless sensor network applications often consisted of homogeneous nodes, current Internet-of-Things/Machine-to-Machine  applications are expected to run on a range of different platforms. A VM can significantly ease the deployment of these applications.

While current VMs offer an impressive set of features, almost all sacrifice performance. The VMs for which we have found concrete performance data are all between one and two orders of magnitude slower than native code. In many scenarios this may not be acceptable for two reasons: for many tasks such as periodic sensing there is a hard limit on the amount of time that can be spent on each measurement, and an application may not be able to tolerate a slowdown of this magnitude. Perhaps more importantly, one of the main reasons for using such tiny devices is their extremely low power consumption. Often, the CPU will be in sleep mode most of the time, so little energy is be spent in the CPU compared to communication, or sensors. But if the slowdown incurred by a VM means the CPU has to stay active 10 to 100 times longer, this may suddenly become the dominant factor.

As an example, one of the few applications reporting a detailed breakdown of its power consumption is Mercury \cite{Lorincz:2009kt}, a platform for motion analysis. The greatest energy consumer is the sampling of a gyroscope, at 53.163 mJ. Only 1.664 mJ is spent in the CPU on application code for an activity recognition filter and feature extraction. When multiplied by 10 or 100 however, the CPU becomes a very significant, or even by far the largest energy consumer. A more complex operation such as a 512 point FFT costs 12.920 mJ. For tasks like this, even a slowdown by a much smaller factor will have a significant impact on the total energy consumption.

A better performing VM is needed, preferably one that performs as close to native performance as possible. Translating bytecode to native code is a common technique to improve performance in desktop VMs. Translation can occur at three moments: offline, ahead-of-time (AOT), or just-in-time (JIT). JIT compilers translate only the necessary parts of bytecode at run-time, just before they are executed. They are common on desktops and on more powerful mobile environments, but are impractical on sensor node platforms that can often only execute code from flash memory. This means a JIT compiler would have to write to flash memory at run-time, which would cause unacceptable delays. Translating to native code offline, before it is sent to the node, has the advantage that more resources are available for the compilation process. We do not have a JVM to AVR compiler to test the resulting performance, but we would expect it would be similar to compiled C code. However, doing so, even if only for small, performance critical sections of code, sacrifices two of the key advantages of using a VM: The host now needs knowledge of the target platform, and needs to prepare a different binary for each type of CPU used in the network, and for the node it will be difficult to provide a safe execution environment when it receives binary code.

Therefore, we focus on the middle option: translating the bytecode to native code on the device itself, at load time. The main research questions to answer are: how close an AOT compiling sensor node VM can come to native C performance, what optimisations are necessary to achieve this, what tradeoffs are involved and what the impact is of the JVM's design decisions for AOT compilation on a sensor node.

\section{Related work}
Many VMs have been proposed that are small enough to fit on a resource-constrained sensor node. They can be divided into two categories: generic VMs and application-specific VMs, or ASVMs \cite{Culler05} that provide specialised instructions for a specific problem domain. One of the first VMs proposed for sensor networks, Mat\'e \cite{Levis:2002ku}, is an ASVM. It provides single instructions for tasks that are common on a sensor node, so programmes can be very short. Unfortunately they have to be written in a low-level assembly-like language, limiting its target audience. SwissQM \cite{Muller:2007fs} is a more traditional VM, based on a subset of the Java VM, but extended with instructions to access sensors and do data aggregation. VM* \cite{Koshy:2005ww} sits halfway between the generic and ASVM approach. It is a Java VM that can be extended with new features according to application requirements. Unfortunately, it is closed source.

Several generic VMs have also been developed, allowing the programmer to use general purpose languages like Java, Python, or even LISP \cite{Harbaum, Brouwers:2009cj, Aslam:2008, Evers:2010}. The smallest official Java standard is the Connected Device Limited Configuration \cite{CLDC}, but since it targets devices with at least a 16 or 32-bit CPU and 160-512KB of flash memory available, it is still too large for most sensor nodes. The available Java VMs for sensor nodes all offer some subset of the standard Java functionality, occupying different points in the tradeoff between the features they provide, and the resources they require.

Only a few papers describing sensor node VMs contain detailed performance measurements. TinyVM \cite{Hong:2009gc} reports a slowdown between 14x and 72x compared to native C, for a set of 9 benchmarks. DVM \cite{Balani:2006} has different versions of the same benchmark, where the fully interpreted version is 108x slower than the fully native version. Ellul reports measurements on the TakaTuka VM \cite{Aslam:2008, Ellul:2012thesis} where the VM is 230x slower than native code, and consumes 150x as much energy. SensorScheme \cite{Evers:2010} is up to 105x slower. Finally, Darjeeling \cite{Brouwers:2009cj} reports between 30x and 113x slowdown. Since performance depends on many factors, it is hard to compare these numbers directly. But the general picture is clear: current interpreters are one to two orders of magnitude slower than native code.

Translating bytecode to native code to improve performance has been a common practice for many years. A wide body of work exists exploring various approaches, either offline, ahead-of-time  or just-in-time. One common offline method is to first translate the Java code to C as an intermediate language, and take advantage of the high quality C compilers available \cite{Muller:1997}. Courbot et al. describe a different approach, where code size is reduced by partly running the application before it is loaded onto the node, allowing them to eliminate code that is only needed during initialisation \cite{Courbot:2010}. Although the initialised objects are translated to C structures that are compiled and linked into a single image, the bytecode is still interpreted. While in general we can produce higher quality code when compiling offline, doing so sacrifices key advantages of using a VM.

Hsieh et al. describe an early ahead-of-time compiling desktop Java VM \cite{Hsieh:1996cy}, focussing on translating the JVM's stack-based architecture to a register based one. In the Japale\~no VM, Alpern et al. take an approach that holds somewhere between AOT and JIT compilation \cite{Alpern:1999}. The VM compiles all code to native code before execution, but can choose from two different compilers to do so. A fast baseline compiler simply mimics the Java stack, but either before or during run-time, a slower optimising compiler may be used to speed up critical methods.

Since JIT compilers work at run-time, much effort has gone into making the compilation process as light weight as possible, for example \cite{Krall:1998}. More recently these efforts have included JIT compilers targeted specifically at embedded devices. Swift \cite{Zhang:2012wf} is a light-weight JVM that improves performance by translating a register-based bytecode to native code. But while the Android devices targeted by Swift may be considered embedded devices, they are still quite powerful and the transformations Swift does are too complex for the ATmega class of devices. HotPathVM \cite{Gal:2006} has lower requirements, but at 150KB for both code and data, this is still an order of magnitude above our target devices.

Given our extreme size constraints - ideally we only want to use in the order of 100 bytes of RAM to allow our approach to be useful on a broad range of devices, and leave ample space for other tasks on the device - almost all AOT and JIT techniques found in literature require too much resources. Indeed, some authors suggest sensor nodes are too restricted to make AOT or JIT compilation feasible \cite{Aslam:2011thesis, Wirjawan:2008}.

On the desktop, VM performance has been studied extensively, but for sensor node VMs this aspect has been mostly ignored. To the best of our knowledge AOT compilation on a sensor node has only been tried by Ellul and Martinez \cite{Ellul:2010iw}, and our work builds on their approach. They improve performance considerably compared to the interpreters, but there is still much room for improvement. Using the standard CoreMark benchmark, their approach generates code that is 811\% slower and 245\% larger than optimised native C. While the reduced throughput may be acceptable for some applications, there are two other reasons why it is important to improve on these results: the loss of performance results in an equivalent increase in cpu power consumption, thus reducing battery life. More importantly, the increased size of the compiled code reduces the amount of code we can load onto a node. Given that flash memory is already restricted, this is a major sacrifice to make when adopting AOT on sensor nodes.

This paper makes the following contributions:
\begin{itemize}
	\item We identify the major sources of overhead when using the baseline approach as described by Ellul and Martinez.
	\item Using the results of this analysis, we propose a set of optimisations to address each source of overhead, including a lightweight alternative to Java method invocation to reduce method call overhead.
	\item These optimisations reduce the code size overhead by 56\%, and show that the increase in VM size is quickly compensated for, thus mitigating a drawback of the previous AOT approach.
	\item They also eliminate most of the performance overhead caused by the JVM's stack-based architecture, and over 80\% of performance overhead overall.
	\item We show that besides these improvements to the AOT technique, better optimisation in the Java to JVM bytecode compiler is critical to achieving good performance.
	\item We provide a comprehensive evaluation to analyse the overhead and the impact of each optimisation, and to show these results hold for a set of benchmarks with very different characteristics, including the commonly used CoreMark benchmark \cite{coremark}.
\end{itemize}

\section{Ahead-of-Time translation}
\label{sec-aot-translation}
Our implementation is based on Darjeeling \cite{Brouwers:2009cj}, a Java VM for sensor nodes, running on an Atmel ATmega CPU. Like other sensor node VMs, it is originally an interpreter. We add an AOT compiler to Darjeeling: instead of interpreting the bytecode, the VM translates it to native code at load time, before the application is started. While JIT compilation is possible on some devices \cite{Ellul:2012thesis}, it depends on the ability to execute code from RAM, which many embedded CPUs, including the ATmega, cannot do.

\begin{figure}[]
  \makebox[\hsize][c]{\includegraphics[width=\linewidth]{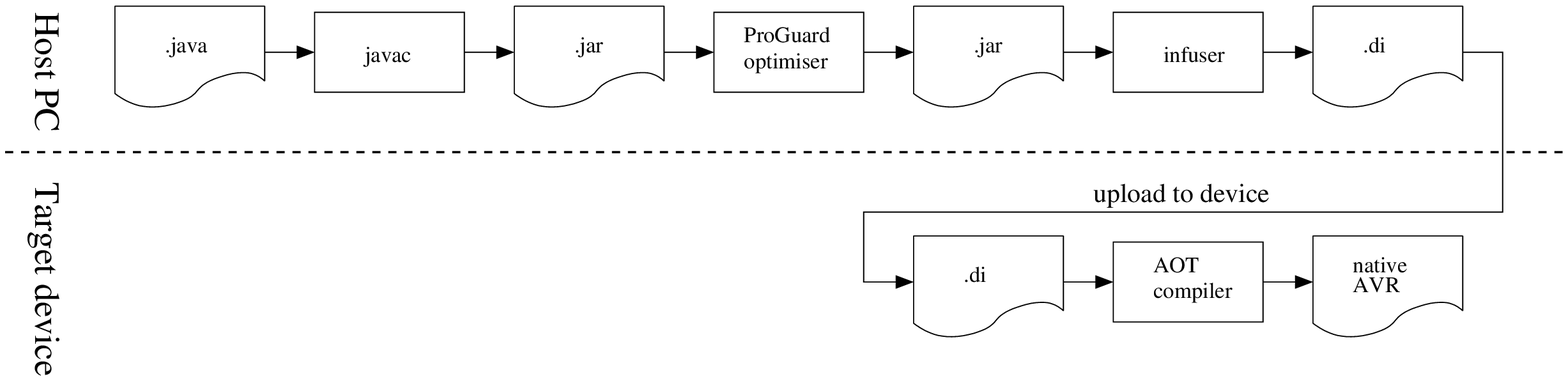}}
  \caption{Java to native AVR compilation}
  \label{fig-translation-process}
\end{figure}

The process from Java source to a native application on the node is shown in Figure \ref{fig-translation-process}. Like all sensor node JVMs, Darjeeling uses a modified JVM bytecode. Java source code is first compiled to normal Java classes, which are optimised by ProGuard \cite{proguard}. The optimised Java classes are then transformed into Darjeeling's own format, called an 'infusion'. For details of this transformation we refer to the Darjeeling paper \cite{Brouwers:2009cj}. Here it is sufficient to know that the bytecode is modified to make it more suitable for execution on a tiny device, for example by adding 16-bit versions of most operations, but the result remains very similar to standard JVM bytecode. It is also important to note that no knowledge of the target platform is used in this transformation, so the result is still platform independent. This infusion is then sent to the node, where it is translated to native AVR code at load time.

We made several modifications to Darjeeling's infuser and bytecode format to support our AOT compiler and improve performance. These changes will be introduced in more detail in the following sections, but for completeness we also list them here:

\begin{itemize}
	\item the \texttt{BRTARGET} opcode, used to mark targets of branch instructions and modified all branch instructions to target a \texttt{BRTARGET} id instead of a bytecode offset
	\item the \texttt{MARKLOOP} opcode to mark inner loops and the variables it uses
	\item added \texttt{\_FIXED} versions of the \texttt{GETFIELD\_A} and \texttt{PUTFIELD\_A} opcodes, used to access an object's reference fields when the offset is known at compile time
	\item the \texttt{SIMUL} opcode for 16x16-bit to 32-bit multiplication
	\item modified array access opcodes to use 16-bit indexes
	\item added \texttt{\_CONST} versions of the bit shift opcodes to support constant shifts
	\item the \texttt{INVOKELIGHT} opcode for an optimised 'lightweight' way of calling methods
\end{itemize}

\subsection{Goals and limitations}
Working on resource-constrained devices means we have to make some compromises. Our main goal is to build a VM that will produce code that both performs well, and adds as little code size overhead as possible. In addition, we want our VM to fit as many scenarios as possible. We would like to be able to support scenarios were multiple applications may be running on a single device, so when new code is being loaded, the impact on other applications should be as small as possible.

Therefore, the translation process should be very light weight. Specifically, it	 should use as little memory as possible, since memory is a very scarce resource. This means we cannot do any analysis on the bytecode that would require us to hold complex data structures in memory. When receiving a large programme, we should not have to keep multiple messages in memory, but will free each message, which can be as small as a single JVM instruction, immediately after processing.

Since messages do need to be processed in the correct order, the actual transmission protocol may still decide to keep more messages in memory to reduce the need for retransmissions in the case of out of order delivery. But our translation process does not require it to do so, and a protocol that values memory usage over retransmissions cost could simply discard out of order messages and request retransmissions when necessary.

Bytecode instructions are processed in a single pass, one instruction at a time. Only some small, fixed-size data structures are kept in memory during the process. A second pass over the generated code then fills in addresses left blank by branch instructions, since the target addresses of forward branches are not known until the target instruction is generated.

The two metrics we compromise on are load time and code size. Compiling to native code takes longer than simply storing bytecode and starting the interpreter, but we feel this load time delay will be acceptable in many cases, and will be quickly compensated for by improved run-time performance. Native code is also larger than JVM bytecode. This is the price we pay for increased performance, but the optimisations we propose do significantly reduce this code size overhead compared to previous work, thus reducing an important drawback of AOT compilation.

Since our compiler is based on Darjeeling, we share its limitations, most notably a lack of floating point support and reflection. In addition, we do not support threads or exceptions because after compilation to native code, we lose the interpreter loop as a convenient place to switch between threads or unwind the stack to jump to an exception handler. Threads and exceptions have been implemented before on a sensor node AOT compiler \cite{Ellul:2012thesis}, proving it is possible to add support for both, but we feel the added complexity in an environment where code space is at a premium makes other, more lightweight models for concurrency and error handling more appropriate.

\subsection{Translating bytecode to native code}
\label{sec-basic-translation}
The basic approach to translate bytecode to native code on a sensor node was first described by Ellul and Martinez \cite{Ellul:2010iw}. When we receive a bytecode instruction, we simply replace it with an equivalent sequence of native instructions, using the native stack to mimic the JVM stack. An example is shown in Table 1.

The first column shows a fragment of JVM code which does a shift right of variable \texttt{A}, and repeats this while \texttt{A} is greater than \texttt{B}. While not a very practical function, it is the smallest example that will allow us to illustrate our code generation optimisations. The second column shows the code the AOT compiler will execute for each JVM instruction. Together, the first and second column match the case labels and body of a big switch statement in our compiler. The third column shows the resulting AVR native code, which is currently almost a 1-on-1 mapping, with the exception of the branch and some small optimisations by a simple peephole optimiser, both described below.

The example has been slightly simplified for readability. Since the AVR is an 8-bit CPU, in the real code many instructions are duplicated for loading the high and low bytes of a short. The cycle count is based on the actual number of generated instructions, and for a single iteration.

\begin{table}[]
\centering
\caption{Translation of \texttt{ do\{A{>}{>}{>}=1;\} while(A>B);}}
\label{tbl-basic-translation}
\small
\makebox[\hsize][c]{\begin{tabular}{llll}
\toprule
JVM & AOT compiler & AVR & cycles \\
\hline
0: BRTARGET(0)   & \sccomment{record current addr} &                &   \\
1: SLOAD\_0      & emit\_LDD(R1,Y+0)        & LDD R1,Y+0     & 4 \\
                 & emit\_PUSH(R1)           & PUSH R1        & 4 \\
2: SCONST\_1     & emit\_LDI(R1,1)          & LDI R1,1       & 2 \\
                 & emit\_PUSH(R1)           & MOV R2,R1      & 1 \\
3: SUSHR         & emit\_POP(R2)            &                &   \\
                 & emit\_POP(R1)            & POP R1         & 4 \\
                 & emit\_RJMP(+2)           & RJMP +2        & 2 \\
                 & emit\_LSR(R1)            & LSR R1         & 2 \\
                 & emit\_DEC(R2)            & DEC R2         & 2 \\
                 & emit\_BRPL(-2)           & BRPL -2        & 3 \\
                 & emit\_PUSH(R1)           &                &   \\
4: SSTORE\_0     & emit\_POP(R1)            &                &   \\
                 & emit\_STD(Y+0,R1)        & STD Y+0,R1     & 4 \\
5: SLOAD\_0      & emit\_LDD(R1,Y+0)        & LDD R1,Y+0     & 4 \\
                 & emit\_PUSH(R1)           & PUSH R1        & 4 \\
6: SLOAD\_1      & emit\_LDD(R1,Y+2)        & LDD R1,Y+2     & 4 \\
                 & emit\_PUSH(R1)           &                &   \\
7: IF\_SCMPGT 0: & emit\_POP(R1)            &                &   \\
                 & emit\_POP(R2)            & POP R2         & 4 \\
                 & emit\_CP(R1,R2)          & CP R1,R2       & 2 \\
                 & emit\_branchtag(GT,0)    & BRGT 0:        & 2 (taken), \\
                 &                          &                & or 1 (not taken) \\
\bottomrule
\end{tabular}}
\end{table}

\subsubsection{Peephole optimisation}
From Table \ref{tbl-basic-translation} it is clear that this approach results in many unnecessary push and pop instructions. Since the JVM is a stack-based VM, each instruction must obtain its operands from the stack and push any result back onto it. As a result, almost half the instructions are push or pop instructions.

To reduce this overhead, Ellul proposes a simple peephole optimiser \cite{Ellul:2012thesis}. The compilation process results in many push instructions that are immediately followed by a pop. If they target the same register, they have no effect and are removed. If the source and destination registers differ, the two instructions are replaced by a move. The result is shown in the third column of Table \ref{tbl-basic-translation}. Two push/pop pairs have been removed, and one has been replaced by a move.

\subsubsection{Branches}
Forward branches pose a problem for our direct translation approach since the target address is not yet known. A second problem is that on the ATmega, a branch may take 1 to 3 words, depending on the distance to the target, so it is also not known how much space should be reserved for a branch.

To solve this the infuser modifies the bytecode by inserting a new instruction, \texttt{BRTARGET}, in front of any instruction that is the target of a branch. The branch instructions themselves are modified to target a branch target id instead of a bytecode offset. When we encounter a \texttt{BRTARGET} during compilation, we do not emit any code, but record the address where the next instruction will be emitted in a separate part of flash. When we encounter a branch instruction, we emit a temporary 3-word 'branch tag' instead, containing the branch target id and the branch condition. After code generation is finished and all target addresses are known, we scan the code again to replace each branch tag with the real branch instruction.

There is still the matter of the different sizes a branch may take. We could simply add \texttt{NOP} instructions to smaller branches to keep the size of each branch at 3 words, but this causes a performance penalty on small, non-taken branches. Instead, we do another scan of the code, before replacing the branch tags, and update the branch target addresses to compensate for cases where a smaller branch will be used. This second scan adds about 500 bytes to the VM, but improves performance, especially on benchmarks where branches are common.

This is an example of something we often see: an optimisation may take a few hundred bytes to implement, but its usefulness may depend on the characteristics of the code being run. In this work we usually decided to implement these optimisations, since they often also result in smaller generated code.

\subsection{Darjeeling split-stack architecture}
\label{sec-darjeeling-split-architecure}
\begin{figure}[]
  \makebox[\hsize][c]{\includegraphics[width=0.7\linewidth]{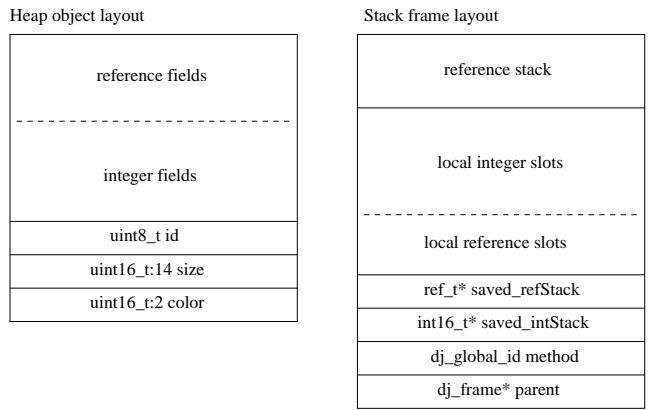}}
  \caption{Object and stack frame layout}
  \label{fig-object-and-stack-frame-layout}
\end{figure}

In Darjeeling, reference and integer values are separated throughout the VM. When the garbage collection runs, it needs to determine which stack values, local variables, static variables and object fields are references. To handle this efficiently, Darjeeling splits references and integers in all these cases, as shown in Figure \ref{fig-object-and-stack-frame-layout}.

In our AOT compiler we use the native stack for the JVM integer operand stack, while space for the reference stack is reserved in the stack frame. This uses less memory than having the integer stack in the stack frame, since we need to reserve space for the maximum stack depth in the frame, which is often much lower for the reference stack than for the integer stack. We use the AVR's X register as a stack pointer for the reference stack.

\subsection{Target platforms}
The AVR family of CPUs is widely used in low power embedded systems. We implemented our VM for the ATmega128 CPU. However, our approach does not depend on any AVR specific properties and we expect similar results for many other CPUs in this class. The main requirements are the ability to reprogramme its own programme memory, and the availability of a sufficient number of registers.

The ATmega128 has 32 8-bit registers. We ran several experiments where we restrict the number of registers our VM may use. As is often the case with caches, we found the first few available registers to have the largest impact, while the added improvement gets less for each added register. Based on this we expect the Cortex M0, with 12 32-bit general purpose registers, or the MSP430, with 12 16-bit registers, and used by Ellul and Martinez \cite{Ellul:2010iw}, to both be good matches as well.

\section{Sources of overhead}
The performance of this basic approach is still far behind optimised native C. To improve performance, it is important to identify the causes of this overhead. The main sources of overhead we found are:
\begin{itemize}
	\item Lack of optimisations in the Java compiler
	\item AOT code generation overhead
	\begin{itemize}
		\item Push/pop overhead
		\item Load/store overhead
		\item JVM instruction set limitations
	\end{itemize}
	\item Method call overhead
\end{itemize}

We will briefly discuss each source of overhead below, before introducing optimisations to reduce it.

\subsection{Lack of optimisation in \texttt{javac}}
A first source of overhead comes from the fact that the standard \texttt{javac} compiler does almost no optimisations.  Since the JVM is an abstract machine, there is no clear performance model to optimise for. Run-time performance depends greatly on the target platform and the VM implementation running the bytecode, which are unknown when compiling Java source code to JVM bytecode.

The \texttt{javac} compiler simply compiles the code 'as is'. For example, the loop '\texttt{while (a < b*c) \{ a*=2; \}}' will evaluate '\texttt{b*c}' on each iteration, while it is clear that the result will be the same every time.

In most environments this is not a problem because the bytecode is typically compiled to native code before execution, and using knowledge of the target platform and the run-time behaviour, a desktop JIT compiler can make much better decisions than \texttt{javac} could. However, since our AOT compiler simply replaces each instruction with a native equivalent, this leads to significant overhead.

We do use the ProGuard optimiser \cite{proguard}, but this only does very basic optimisations such as method inlining and dead code removal, and does not cover cases such as the example above.

\subsection{AOT translation overhead}
\label{sec-overhead-aot-translation}
Assuming we have high quality JVM bytecode, a second source of overhead comes from the way the bytecode is translated to native code. We distinguish three main types of translation overhead, where the first two are a direct result of the JVM's stack-based architecture.

\subsubsection{Type 1: Pushing and popping values} The compilation process initially results in a large number of push and pop instructions. In our simple example in Table \ref{tbl-basic-translation}, the peephole optimiser was able to eliminate some, but two push/pop pairs remain. For more complex expressions this type of overhead is even higher, since more values will be on the stack at the same time. This means more corresponding push and pop instructions will not be consecutive, and the peephole optimiser cannot eliminate these cases.

\subsubsection{Type 2: Loading and storing values} The second type is also due to the JVM's stack-based architecture. Each operation consumes its operands from the stack, but often the same value is needed again soon after. In this case, because the value is no longer on the stack, we need to do another load, which will result in another read from memory.

In Table \ref{tbl-basic-translation}, it is clear that the \texttt{SLOAD\_0} instruction at label 5 is unnecessary since the value is already in R1.

\subsubsection{Type 3: JVM instruction set limitations} A final source of overhead comes from optimisations that are done in native code, but are not possible in JVM bytecode, at least not in our resource-constrained environment.

The JVM instruction set is very simple, which makes it easy to implement, but this also means some things cannot be expressed as efficiently as in native code. Given enough processing power, compilers can do the complex transformations necessary to make the compiled JVM code run almost as fast as native C, but on a sensor node we do not have such resources and must simply execute the instructions as they are.

In Table \ref{tbl-basic-translation} we see that there is no way to express a single bit shift directly. Instead we have to load the constant 1 onto the stack and execute the generic bit shift instruction. Compare this to addition, where the JVM bytecode does have a special INC instruction to add a constant value to a local variable.

A second example is array access. In JVM bytecode each array access will consume the array reference and index from the stack. When looping over an array, this means we that for each iteration we have to load the reference and index back onto the stack again, and redo the address calculation. In contrast, the native C version would typically just slide a pointer over the array.

\subsection{Method call overhead}
\label{sec-overhead-method-call}
The final source of overhead comes from method calls. In the JVM, each method has a stack frame (or 'activation frame') which the language specification describes as
\begin{quotation}
"containing the target reference (if any) and the argument values (if any), as well as enough space for the local variables and stack for the method to be invoked and any other bookkeeping information that may be required by the implementation (stack pointer, program counter, reference to previous activation frame, and the like)" \cite{Gosling:2014}
\end{quotation}

Darjeeling's stack frame layout is shown in Figure \ref{fig-object-and-stack-frame-layout}. Initialising this complete structure is significantly more work than a native C function call has to do, which may not need a stack frame at all if all the work can be done in registers. Below we list the steps Darjeeling goes through to invoke a Java method:

\begin{enumerate}
  \small
  \item flush the stack cache so parameters are in memory and clear value tags (see sections \ref{sec-optimisations-simple-stack-caching} and \ref{sec-optimisations-popped-value-caching})
  \item save int and ref stack pointers (SP and X)
  \item call the VM's \texttt{callMethod} function, which will:
  \begin{enumerate}
    \item allocate memory for the callee's frame
    \item initialise the callee's frame
    \item pass parameters: pop them off the caller's stack and copy them into the callee's locals
    \item activate the callee's frame: set the VM's active frame pointer to the callee
    \item lookup the address of the AOT compiled code
    \item do the actual \texttt{CALL}, which will return any return value in registers R22 and higher
    \item reactivate the old frame: set the VM's active frame pointer back to the caller
    \item return to the caller's AOT compiled code the return value (if any) in R22 and higher
  \end{enumerate}
  \item restore stack pointer and X register
  \item push the return value onto the stack (using stack caching, so this is free)
\end{enumerate}

Even after considerable effort optimising this process, this requires roughly 550 cycles for the simplest case: a call to a static method without any parameters or return value. For a virtual method the cost is higher because we need to look up the right implementation. While we may be able to save some more cycles with an even more rigorous refactoring, it is clear that the number of steps involved will always take considerably more time than a native function call.

\subsection{Optimisations}
\label{sec-optimisations-java-source}
Having identified these sources of overhead, we will use the next three sections to describe the set of optimisations we use to address them. Table \ref{tbl-optimisations-overview} lists each optimisation, and the source of overhead it aims to reduce. The following sections will discuss each optimisation in detail.

\begin{table*}[hbt]
\centering
\caption{List of optimisations per overhead source}
\label{tbl-optimisations-overview}
\small
\makebox[\hsize][c]{\begin{tabular}{lll}
\toprule
& Source of overhead & Optimisation \\
\hline
Section \ref{sec-optimisations-manual-java-source-optimisation} &
Lack of optimisations in \texttt{javac}        & Manual optimisation of Java source code \\

Section \ref{sec-optimisations-aot-translation-overhead} &
AOT translation overhead                       & \\
&\hspace{.5cm} Push/pop overhead                & Improved peephole optimiser \\
&                                               & Stack caching \\
&\hspace{.5cm} Load/store overhead              & Popped value caching \\
&                                               & Mark loops \\
&\hspace{.5cm} JVM instruction set limitations  & \texttt{SIMUL} instruction \\
&                                               & \texttt{GET/PUTFIELD\_A\_FIXED} instructions \\
&                                               & constant shift optimisation \\
&                                               & 16-bit array indexes \\

Section \ref{sec-optimisations-method-calls} &
Method call overhead                           & \texttt{INVOKELIGHT} instruction \\
\bottomrule
\end{tabular}}
\end{table*}

\section{Manually optimising the Java source code}
\label{sec-optimisations-manual-java-source-optimisation}
As shown in Section \ref{sec-aot-translation}, our current implementation uses three steps to translate Java source code to Darjeeling bytecode: the standard Java compiler, the ProGuard optimiser, and Darjeeling's infuser. None of these do any complex optimisations. 

In a future version, ProGuard and the infuser should be merged into an 'optimising infuser' which uses normal, well-known optimisation techniques to produce better quality bytecode.

At the moment we do not have the resources to build such an optimising infuser. Since our goal is to find out what level of performance is possible on a sensor node, we manually optimise the Java source to get better quality JVM bytecode from \texttt{javac}. While these changes are not an automatic optimisation we developed, we find it imporant to mention them explicitly and analyse their impact, since many developers may expect many of these to happen automatically, and without this it would be impossible to reproduce our results.

We have been careful to limit ourselves to 'fair' optimisations, by which we mean optimisations that an optimising infuser could reasonably be expected to do automatically, given some basic, conservative assumptions about the performance model. 

The most common optimisations we performed are:
\begin{itemize}
	\item store the result of expressions calculated in a loop in a temporary variable, if it is known the result will be the same for each iteration
	\item since array and object field access is relatively expensive and not cached by the mark loop optimisation discussed in Section \ref{sec-optimisation-markloops}, prefer to store a value in a local variable if it may be used again soon rather than accessing the array or object twice
	\item manually inlining small methods
	\item prefer to use 16-bit variables for array indexes where possible
	\item use bit shifts for multiplications by a power of two
\end{itemize}

We will briefly examine the effect of some 'unfair' optimisations on the Core Mark benchmark in Section \ref{sec-evaluation-coremark}.

\paragraph{Manual inlining}
We manually inline all small methods that were either a \texttt{\#define} in the original C code, or a function that was inlined by \texttt{avr-gcc}. ProGuard can also inline small methods, but when it does, it simply replaces the \texttt{INVOKE} instruction with the callee's body, prepended with \texttt{STORE} instructions to pop the parameters off the stack and initialise the callee's local variables. Manual inlining often results in better code, because it may not be necessary to store the parameters if they are only used once. Again, it is easy to imagine that an optimising compiler should be able to come to the same result automaticallly.

\paragraph{Platform independence}
Assuming an optimising infuser does raise the question how platform independent the resulting code is. If the infuser has more specific knowledge about the target platform, it can produce better code for that platform, but, while it should still run anywhere, this may not be as efficient on other platforms.

However, the optimisations described here are only based on very conservative assumptions that would work well for most devices in this class.

\paragraph{Example} An example of these manual optimisations, applied to the bubble sort benchmark, can be seen in Listing \ref{lst-manual-optimisation}. To have a fair comparison, we applied exactly the same optimisations to the C versions of our benchmarks, but here this had little or no effect on the performance.

\begin{listing}[H]
 \centering
 \begin{minipage}[t]{0.45\textwidth}
  \centering
  \begin{minted}[fontsize=\scriptsize]{java}
// ORIGINAL
public static void bsort(int[] numbers) {
  short NUMNUMBERS=(short)numbers.length;
  for (short i=0; i<NUMNUMBERS; i++) {
    for (short j=0; j<NUMNUMBERS-i-1; j++) {
      if (numbers[j]>numbers[j+1]) {
        int temp = numbers[j];
        numbers[j] = numbers[j+1];
        numbers[j+1] = temp;
      }
    }
  }
}
  \end{minted}
 \end{minipage}\hfill
 \begin{minipage}[t]{0.45\textwidth}
  \centering
  \begin{minted}[fontsize=\scriptsize]{java}
// MANUALLY OPTIMISED
public static void bsort(int[] numbers) {
  short NUMNUMBERS=(short)numbers.length;
  for (short i=0; i<NUMNUMBERS; i++) {
    short x=(short)(NUMNUMBERS-i-1);
    short j_plus_one = 1;
    for (short j=0; j<x; j++) {
      int val_at_j = numbers[j];
      int val_at_j_plus_one = numbers[j_plus_one];
      if (val_at_j>val_at_j_plus_one) {
        numbers[j] = val_at_j_plus_one;
        numbers[j_plus_one] = val_at_j;
      }
      j_plus_one++;
    }
  }
}
  \end{minted}
 \end{minipage}
\caption{Optimisation of the bubble sort benchmark}
\label{lst-manual-optimisation}
\end{listing}

\section{Optimisations: AOT translation overhead}
\label{sec-optimisations-aot-translation-overhead}
Now that we have good quality bytecode to work with, we can start addressing the overhead incurred during the AOT compilation process.

\subsection{Improving the peephole optimiser}
\label{sec-improved-peephole}

Our first optimisation is a small but effective extension to the simple peephole optimiser. Instead of optimising only consecutive push/pop pairs, we can optimise any pair of push/pop instructions if the following holds for the instructions in between:

\begin{listing}[H]
 \centering
 \begin{minted}[fontsize=\scriptsize]{asm}
PUSH Rs
..
..       instructions in between: - contain the same number of push and pop instructions
..                                - contain no branches
..                                - do not use register Rd
..
POP  Rd
 \end{minted}
\end{listing}

In this case the pair can be eliminated if \texttt{Rs} == \texttt{Rd}, otherwise it is replaced by a '\texttt{mov Rd, Rs}'. Two push/pop pairs remain in Table \ref{tbl-basic-translation}. The pair in instructions 5 and 7 pops to register R2. Since instruction 6 does not use register R2, we can safely replace this pair with a direct move. In contrast, the pair in instructions 1 and 3 cannot be optimised since the value is popped into register R1, which is also used by instruction 2.

\subsection{Simple stack caching}
\label{sec-optimisations-simple-stack-caching}
\begin{table*}[hbt]
\centering
\caption{Simple stack caching}
\label{tbl-simplestackcaching}
\scriptsize
\addtolength{\tabcolsep}{-2pt}
\makebox[\hsize][c]{\begin{tabular}{llll|c|c|c}
\toprule
JVM                & AOT compiler                                         & AVR                 & cycles & cache state R1                   & cache state R2                   & cache state R3                   \\
\hline
0: BRTARGET(0)     & \sccomment{record current addr} & & & & & \\
1: SLOAD\_0        & operand\_1 = sc\_getfreereg()                        &                     &        & \stackcacheentry{    }{   }{   } & \stackcacheentry{    }{   }{   } & \stackcacheentry{    }{   }{   } \\
                   & emit\_LDD(operand\_1,Y+0)                            & LDD R1,Y+0          & 4      & \stackcacheentry{    }{   }{   } & \stackcacheentry{    }{   }{   } & \stackcacheentry{    }{   }{   } \\
                   & sc\_push(operand\_1)                                 &                     &        & \stackcacheentry{Int1}{   }{   } & \stackcacheentry{    }{   }{   } & \stackcacheentry{    }{   }{   } \\
3: SUSHR\_CONST(1) & operand\_1 = sc\_pop()                               &                     &        & \stackcacheentry{    }{   }{   } & \stackcacheentry{    }{   }{   } & \stackcacheentry{    }{   }{   } \\
                   & emit\_LSR(operand\_1)                                & LSR R1              & 2      & \stackcacheentry{    }{   }{   } & \stackcacheentry{    }{   }{   } & \stackcacheentry{    }{   }{   } \\
                   & sc\_push(operand\_1)                                 &                     &        & \stackcacheentry{Int1}{   }{   } & \stackcacheentry{    }{   }{   } & \stackcacheentry{    }{   }{   } \\
4: SSTORE\_0       & operand\_1 = sc\_pop()                               &                     &        & \stackcacheentry{    }{   }{   } & \stackcacheentry{    }{   }{   } & \stackcacheentry{    }{   }{   } \\
                   & emit\_STD(Y+0,operand\_1)                            & STD Y+0,R1          & 4      & \stackcacheentry{    }{   }{   } & \stackcacheentry{    }{   }{   } & \stackcacheentry{    }{   }{   } \\
5: SLOAD\_0        & operand\_1 = sc\_getfreereg()                        &                     &        & \stackcacheentry{    }{   }{   } & \stackcacheentry{    }{   }{   } & \stackcacheentry{    }{   }{   } \\
                   & emit\_LDD(operand\_1,Y+0)                            & LDD R1,Y+0          & 4      & \stackcacheentry{    }{   }{   } & \stackcacheentry{    }{   }{   } & \stackcacheentry{    }{   }{   } \\
                   & sc\_push(operand\_1)                                 &                     &        & \stackcacheentry{Int1}{   }{   } & \stackcacheentry{    }{   }{   } & \stackcacheentry{    }{   }{   } \\
6: SLOAD\_1        & operand\_1 = sc\_getfreereg()                        &                     &        & \stackcacheentry{Int1}{   }{   } & \stackcacheentry{    }{   }{   } & \stackcacheentry{    }{   }{   } \\
                   & emit\_LDD(operand\_1,Y+2)                            & LDD R2,Y+2          & 4      & \stackcacheentry{Int1}{   }{   } & \stackcacheentry{    }{   }{   } & \stackcacheentry{    }{   }{   } \\
                   & sc\_push(operand\_1)                                 &                     &        & \stackcacheentry{Int2}{   }{   } & \stackcacheentry{Int1}{   }{   } & \stackcacheentry{    }{   }{   } \\
7: IF\_SCMPGT 0:   & operand\_1 = sc\_pop()                               &                     &        & \stackcacheentry{Int1}{   }{   } & \stackcacheentry{    }{   }{   } & \stackcacheentry{    }{   }{   } \\
                   & operand\_2 = sc\_pop()                               &                     &        & \stackcacheentry{    }{   }{   } & \stackcacheentry{    }{   }{   } & \stackcacheentry{    }{   }{   } \\
                   & emit\_CP(operand\_1, operand\_2);                    & CP R2,R1            & 2      & \stackcacheentry{    }{   }{   } & \stackcacheentry{    }{   }{   } & \stackcacheentry{    }{   }{   } \\
                   & emit\_branchtag(GT, 0);                              & BRGT 0:             & 2 or 1 & \stackcacheentry{    }{   }{   } & \stackcacheentry{    }{   }{   } & \stackcacheentry{    }{   }{   } \\
\bottomrule
\end{tabular}}
\addtolength{\tabcolsep}{2pt}
\end{table*}

The improved peephole optimiser can remove part of the type 1 overhead, but still many cases remain where it cannot eliminate the push/pop instructions. We use a form of stack caching \cite{Ertl:1995dv} to eliminate most of the remaining push/pop overhead. Stack caching is not a new technique, but the tradeoffs are very different depending on the scenario it is applied in, and it turns out to be exceptionally well suited for a sensor node AOT compiler:

First, the VM in the original paper is an interpreter, which means the stack cache has to be very lightweight, otherwise the overhead from managing it at run-time will outweigh the time saved by reducing memory accesses. Since we only use the cache state at load time, this restriction does not apply for an AOT compiler and we can afford to spend more time managing the cache. Second, the simplicity of the approach means it requires very little memory: only 11 bytes of RAM and less than 1KB of code more than the peephole optimiser.

The basic idea of stack caching is to keep the top elements of the stack in registers instead of main memory. We add a cache state to our VM to keep track of which registers are holding stack elements. For example, if the top two elements are kept in registers, an ADD instruction does not need to access main memory, but can simply add these registers, and update the cache state. Values are only spilled to memory when all registers available for stack caching are in use.

In the original approach, each JVM instruction maps to a fixed sequence of native instructions that always use the same registers. Using stack caching, the registers are controlled by a stack cache manager that provides three functions:
\begin{itemize}
    \item \texttt{getfree}: Instructions such as load instructions will need a free register to load the value into, which will later be pushed onto the stack. If all registers are in use, \texttt{getfree} spills the register that's lowest on the stack to memory by emitting a \texttt{PUSH}, and then returns that register. This way the top of the stack is kept in registers, while lower elements may be spilled to memory.
    \item \texttt{pop}: Pops the top element off the stack and tells the code generator in which register to find it. If stack elements have previously been spilled to main memory and no  elements are left in registers, \texttt{pop} will emit a real \texttt{POP} instruction to get the value back from memory.
    \item \texttt{push}: Updates the cache state so the passed register is now at the top of the stack. This should be a register that was previously returned by \texttt{getfree}, or \texttt{pop}.
\end{itemize}

Using stack caching, code generation is split between the instruction translator, which emits the instructions that do the actual work, and the cache manager which manages the registers and may emit code to spill stack elements to memory, or to retrieve them again. But as long as enough registers are available, it will only manipulate the cache state.

In Table \ref{tbl-simplestackcaching} we translate the same example we used before, but this time using stack caching. To save space, Table \ref{tbl-simplestackcaching} also includes the constant shift optimisation described in Section \ref{sec-opt-constant-shift}. The \texttt{emit\_PUSH} and \texttt{emit\_POP} instructions have been replaced by calls to the cache manager, and instructions that load something onto the stack start by asking the cache manager for a free register. The state of the stack cache is shown in the three columns added to the right. Currently it only tracks whether a register is on the stack or not. "Int1" marks the top element, followed by "Int2", etc. (this example does not use the reference stack) In the next two optimisations we will extend the cache state further.
 
The example only shows three registers, but the ATmega128 we use has 32 8-bit registers. Since Darjeeling uses a 16-bit stack, we manage them as pairs. 10 registers are reserved, for example as a scratch register or to store a pointer to local or static variables, leaving 11 pairs available for stack caching.

\paragraph{Branches} Branch targets may be reached from multiple locations. We know the cache state if it was reached from the previous instruction, but not if it was reached through a branch. To ensure the cache state is the same on both paths, we flush the whole stack to memory whenever we encounter either a branch or a \texttt{BRTARGET} instruction. 

This may seem bad for performance, but fortunately in the code generated by \texttt{javac} the stack is empty at almost all branches. The exception is the ternary \texttt{?} \texttt{:} operator, which may cause a conditional branch with elements on the stack, but in most cases flushing at branches and branch targets does not result in any extra overhead.

\subsection{Popped value caching}
\label{sec-optimisations-popped-value-caching}
Stack caching can eliminate most of the push/pop overhead, even when the stack depth increases. We now turn our attention to reducing the overhead resulting from load and store instructions.

\begin{table*}[hbt]
\centering
\caption{Popped value caching}
\label{tbl-poppedvaluecaching}
\scriptsize
\addtolength{\tabcolsep}{-2pt}
\makebox[\hsize][c]{\begin{tabular}{llll|c|c|c}
\toprule
JVM                & AOT compiler                                         & AVR                 & cycles & cache state R1                   & cache state R2                   & cache state R3                   \\
\hline
0: BRTARGET(0)   & \sccomment{record current addr} & & & & & \\
1: SLOAD\_0        & operand\_1 = sc\_getfreereg()                        &                     &        & \stackcacheentry{    }{   }{   } & \stackcacheentry{    }{   }{   } & \stackcacheentry{    }{   }{   } \\
                   & emit\_LDD(operand\_1,Y+0)                            & LDD R1,Y+0          & 4      & \stackcacheentry{    }{   }{   } & \stackcacheentry{    }{   }{   } & \stackcacheentry{    }{   }{   } \\
                   & sc\_push(operand\_1)                                 &                     &        & \stackcacheentry{Int1}{LS0}{   } & \stackcacheentry{    }{   }{   } & \stackcacheentry{    }{   }{   } \\
3: SUSHR\_CONST(1) & operand\_1 = sc\_pop\_destructive()                  &                     &        & \stackcacheentry{    }{   }{   } & \stackcacheentry{    }{   }{   } & \stackcacheentry{    }{   }{   } \\
                   & emit\_LSR(operand\_1)                                & LSR R1              & 2      & \stackcacheentry{    }{   }{   } & \stackcacheentry{    }{   }{   } & \stackcacheentry{    }{   }{   } \\
                   & sc\_push(operand\_1)                                 &                     &        & \stackcacheentry{Int1}{   }{   } & \stackcacheentry{    }{   }{   } & \stackcacheentry{    }{   }{   } \\
4: SSTORE\_0       & operand\_1 = sc\_pop\_tostore()                      &                     &        & \stackcacheentry{    }{LS0}{   } & \stackcacheentry{    }{   }{   } & \stackcacheentry{    }{   }{   } \\
                   & emit\_STD(Y+0,operand\_1)                            & STD Y+0,R1          & 4      & \stackcacheentry{    }{LS0}{   } & \stackcacheentry{    }{   }{   } & \stackcacheentry{    }{   }{   } \\
5: SLOAD\_0        & \sccomment{skip codegen, just update cache state}    &                     &        & \stackcacheentry{Int1}{LS0}{   } & \stackcacheentry{    }{   }{   } & \stackcacheentry{    }{   }{   } \\
6: SLOAD\_1        & operand\_1 = sc\_getfreereg()                        &                     &        & \stackcacheentry{Int1}{LS0}{   } & \stackcacheentry{    }{   }{   } & \stackcacheentry{    }{   }{   } \\
                   & emit\_LDD(operand\_1,Y+2)                            & LDD R2,Y+2          & 4      & \stackcacheentry{Int1}{LS0}{   } & \stackcacheentry{    }{   }{   } & \stackcacheentry{    }{   }{   } \\
                   & sc\_push(operand\_1)                                 &                     &        & \stackcacheentry{Int2}{LS0}{   } & \stackcacheentry{Int1}{LS1}{   } & \stackcacheentry{    }{   }{   } \\
7: IF\_SCMPGT 0:   & operand\_1 = sc\_pop\_nondestructive()               &                     &        & \stackcacheentry{Int1}{LS0}{   } & \stackcacheentry{    }{LS1}{   } & \stackcacheentry{    }{   }{   } \\
                   & operand\_2 = sc\_pop\_nondestructive()               &                     &        & \stackcacheentry{    }{LS0}{   } & \stackcacheentry{    }{LS1}{   } & \stackcacheentry{    }{   }{   } \\
                   & emit\_CP(operand\_1, operand\_2);                    & CP R2,R1            & 2      & \stackcacheentry{    }{LS0}{   } & \stackcacheentry{    }{LS1}{   } & \stackcacheentry{    }{   }{   } \\
                   & emit\_branchtag(GT, 0);                              & BRGT 0:             & 2 or 1 & \stackcacheentry{    }{LS0}{   } & \stackcacheentry{    }{LS1}{   } & \stackcacheentry{    }{   }{   } \\
\bottomrule
\end{tabular}}
\addtolength{\tabcolsep}{2pt}
\end{table*}

We add a 'value tag' to each register's cache state to keep track of what value is currently held in the register, even after it is popped from the stack. Some JVM instructions have a value tag associated with them to indicate which value or variable they load, store, or modify. Each tag consist of a tuple (type, datatype, number). For example, the JVM instructions \texttt{ILOAD\_0} and \texttt{ISTORE\_0}, which load and store the local integer variable with id 0, both have tag LI0, short for (local, int, 0). \texttt{SCONST\_1} has tag CS1, or (constant, short, 1), etc. These tags are encoded in a 16-bit value.

We add a function, \texttt{sc\_can\_skip}, to the cache manager. This function will examine the type of each instruction, its value tag, and the cache state. If it finds that we are loading a value that is already present in a register, it updates the cache state to put that register on the stack, and returns true to tell the main loop to skip code generation for this instruction.

Table \ref{tbl-poppedvaluecaching} shows popped value caching applied to our example. At first, the stack is empty. When \texttt{sc\_push} is called, it detects the current instruction's value tag, and marks the fact that R1 now contains LS0. In \texttt{SUSHR\_CONST}, the \texttt{pop} has been changed to \texttt{pop\_destructive}. This tells the cache manager that the value in the register will be destroyed, so the value tag has to be cleared again since R1 will no longer contain LS0. The \texttt{SSTORE\_0} instruction now calls \texttt{pop\_tostore} instead of  \texttt{pop}, to inform the cache manager it will store this value in the variable identified by \texttt{SSTORE\_0}'s value tag. This means the register once again contains LS0. If any other register was marked as containing LS0, the cache manager would clear that tag, since it is no longer accurate after we update the variable.

In line 5, we need to load LS0 again, but now the cache state shows that LS0 is already in R1. This means we do not need to load it from memory, but just update the cache state so that R1 is pushed onto the stack. At run-time this \texttt{SLOAD\_0} will have no cost at all.

There are a few more details to get right. For example if we load a value that's already on the stack, we generate a move to copy it. When \texttt{sc\_getfree} is called, it will try to return a register without a value tag. If none are available, the least recently used register is returned. This is done to maximise the chance we can reuse a value later, since recently used values are more likely to be used again.

\paragraph{Branches} As we do not know the state of the registers if an instruction is reached through a branch, we have to clear all value tags when we pass a \texttt{BRTARGET} instruction, meaning that any new loads will have to come from memory. At branches we can keep the value tags, because if the branch is not taken, we do know the state of the registers in the next instruction.

\subsection{Mark loops}
\label{sec-optimisation-markloops}
\begin{table*}[hbt]
\centering
\caption{Mark loops}
\label{tbl-markloop}
\scriptsize
\addtolength{\tabcolsep}{-2pt}
\makebox[\hsize][c]{\begin{tabular}{llll|c|c|c}
\toprule
JVM                & AOT compiler                                            & AVR                 & cycles & cache state R1                      & cache state R2                      & cache state R3                   \\
\hline
0: MARKLOOP(0,1)   & \emph{ << emit markloop prologue:}                      & LDD R1,Y+0          & 4      & \stackcacheentry{    }{LS0}{PIN   } & \stackcacheentry{    }{   }{      } & \stackcacheentry{    }{   }{   } \\
                   &  \emph{\hspace{.7cm}LS0 and LS1 are live >> }           & LDD R2,Y+2          & 4      & \stackcacheentry{    }{LS0}{PIN   } & \stackcacheentry{    }{LS1}{PIN   } & \stackcacheentry{    }{   }{   } \\
1: BRTARGET(0)     & \sccomment{record current addr}                         &                     &        & \stackcacheentry{    }{LS0}{PIN   } & \stackcacheentry{    }{LS1}{PIN   } &                                  \\
2: SLOAD\_0        & \sccomment{skip codegen, just update cache state}       &                     &        & \stackcacheentry{Int1}{LS0}{PIN   } & \stackcacheentry{    }{LS1}{PIN   } & \stackcacheentry{    }{   }{   } \\
4: SUSHR\_CONST(1) & operand\_1 = sc\_pop\_destructive()                     & MOV R3,R1           & 1      & \stackcacheentry{    }{LS0}{PIN   } & \stackcacheentry{    }{LS1}{PIN   } & \stackcacheentry{    }{   }{   } \\
                   & emit\_LSR(operand\_1)                                   & LSR R3              & 2      & \stackcacheentry{    }{LS0}{PIN   } & \stackcacheentry{    }{LS1}{PIN   } & \stackcacheentry{    }{   }{   } \\
                   & sc\_push(operand\_1)                                    &                     &        & \stackcacheentry{    }{LS0}{PIN   } & \stackcacheentry{    }{LS1}{PIN   } & \stackcacheentry{Int1}{   }{   } \\
5: SSTORE\_0       & \sccomment{skip codegen, move to pinned reg}            & MOV R1,R3           & 1      & \stackcacheentry{    }{LS0}{PIN   } & \stackcacheentry{    }{LS1}{PIN   } & \stackcacheentry{    }{   }{   } \\
6: SLOAD\_0        & \sccomment{skip codegen, just update cache state}       &                     &        & \stackcacheentry{Int1}{LS0}{PIN   } & \stackcacheentry{    }{LS1}{PIN   } & \stackcacheentry{    }{   }{   } \\
7: SLOAD\_1        & \sccomment{skip codegen, just update cache state}       &                     &        & \stackcacheentry{Int2}{LS0}{PIN   } & \stackcacheentry{Int1}{LS1}{PIN   } & \stackcacheentry{    }{   }{   } \\
8: IF\_SCMPGT 0:   & operand\_1 = sc\_pop\_nondestructive()                  &                     &        & \stackcacheentry{Int1}{LS0}{PIN   } & \stackcacheentry{    }{LS1}{PIN   } & \stackcacheentry{    }{   }{   } \\
                   & operand\_2 = sc\_pop\_nondestructive()                  &                     &        & \stackcacheentry{    }{LS0}{PIN   } & \stackcacheentry{    }{LS1}{PIN   } & \stackcacheentry{    }{   }{   } \\
                   & emit\_CP(operand\_1, operand\_2);                       & CP R2,R1            & 2      & \stackcacheentry{    }{LS0}{PIN   } & \stackcacheentry{    }{LS1}{PIN   } & \stackcacheentry{    }{   }{   } \\
                   & emit\_branchtag(GT, 0);                                 & BRGT 1:             & 2 or 1 & \stackcacheentry{    }{LS0}{PIN   } & \stackcacheentry{    }{LS1}{PIN   } & \stackcacheentry{    }{   }{   } \\
9: MARKLOOP(end)   & \sccomment{emit markloop epilogue: LS0 is live}         & STD Y+0,R1          & 4      & \stackcacheentry{    }{LS0}{      } & \stackcacheentry{    }{LS1}{      } & \stackcacheentry{    }{   }{   } \\
\bottomrule
\end{tabular}}
\addtolength{\tabcolsep}{2pt}
\end{table*}

Popped value caching reduces the type 2 overhead significantly, but the fact that we have to clear the value tags at branch targets means that a large part of that overhead still remains. This is particularly true for loops, since each iteration often uses the same variables, but the branch to start the next iteration clears those values from the stack cache. This is addressed by the next optimisation.

Again, we modify the infuser to add a new instruction to the bytecode: \texttt{MARKLOOP}. This instruction is used to mark the beginning and end of each inner loop. \texttt{MARKLOOP} has a larger payload than most JVM instructions: it contains a list of value tags that will appear in the loop and how often each tag appears, sorted in descending order.

When we encounter the \texttt{MARKLOOP} instruction, the VM may decide to reserve a number of registers and pin the most frequently used local variables to them. If it does, code is generated to prefetch these variables from memory and store them in registers. While in the loop, loading or storing these pinned variables does not require memory access, but only a manipulation of the cache state, and possibly a simple move between registers. However, these registers will no longer be available for normal stack caching. Since 4 register pairs need to be reserved for code generation, at most 7 of the 11 available pairs can be used by mark loops.

Because the only way to enter and leave the loop is through the \texttt{MARKLOOP} instructions, the values can remain pinned for the whole duration of the block, regardless of the branches made inside. This lets us eliminate more load instructions, and also replace store instructions by a much cheaper move to the pinned register. \texttt{INC} instructions, which increment a local variable, operate directly on the pinned register, saving both a load and a store. All these cases are handled in \texttt{sc\_can\_skip}, bypassing the normal code generation. We also need to make a small change to \texttt{sc\_pop\_destructive}. If the register we're about to pop is pinned, we cannot just return it since it would corrupt the value of the pinned local variable. Instead we will first emit a move to a free, non-pinned register, and return that instead.

In Table \ref{tbl-markloop} the first instruction is now \texttt{MARKLOOP}, which tells the compiler local short variables 0 and 1 will be used. The compiler decides to pin them both to registers 1 and 2. The \texttt{MARKLOOP} instruction also tells the VM whether or not the variables are live, which they are at this point, so the two necessary loads are generated. This is reflected in the cache state. No elements are on the stack yet, but register 1 is pinned to LS0, and register 2 to LS1.

Next, LS0 is loaded. Since it is pinned to register 1, no code is generated, but the cache state is updated to reflect LS0 is now on top of the stack. Next, \texttt{SUSHR\_CONST} pops destructively. We cannot simply return register 1 since that would corrupt the value of variable LS0, so \texttt{sc\_pop\_destructive} emits a move to a free register and returns that register instead. Since LS0 is pinned, we can also skip \texttt{SSTORE\_0}, but we do need to emit a move back to the pinned register.

The next two loads are straightforward and can be skipped, and in the branch we see the registers are popped non-destructively, so we can use the pinned registers directly.

Finally, we see the loop ends with another \texttt{MARKLOOP}, telling the compiler only local 0 is live at this point. This means we need to store LS0 in register 1 back to memory, but we can skip LS1 since it is no longer needed.

The total cost is now 20 cycles, which appears to be up two from the 18 cycles spent using only popped value caching. But 12 of these are spent before and after the loop, while each iteration now only takes 8 cycles, a significant improvement from the 48 cycles spent in the original version in Table \ref{tbl-basic-translation}.

\subsection{Instruction set modifications}
Next, we introduce four optimisations that target the type 3 overhead: cases where limitations in the JVM instruction set means we cannot express some operations as efficiently as we would like. This type of overhead is the most difficult to address because many of the transformations a desktop VM can do to avoid it take more resources than we can afford on a tiny device. Also, this type of overhead covers many different cases, and optimisations that help in a specific case may not be general enough to justify spending additional resources on it.

Still, there are a few things we can do by modifying the instruction set, that come at little cost to the VM and can make a significant difference.

Darjeeling's original instruction set is already quite different from the normal JVM instruction set. The most important change is the introduction of 16-bit operations. The JVM is internally a 32-bit machine, meaning \texttt{short}, \texttt{byte}, and \texttt{char} are internally stored as 32-bit integers. On a sensor device where memory is the most scarce resource, we often want to use shorter data types. To support this, Darjeeling internally stores values in 16-bit slots, and introduces 16-bit versions of all integer operations. For example if we want to multiply two shorts and store the result in a short, the 32-bit \texttt{IMUL} instruction is replaced by the 16-bit \texttt{SMUL} instruction. These transformations are all done by the infuser (see Figure \ref{fig-translation-process}).

However, the changes made by Darjeeling are primarily aimed at reducing memory consumption, not at improving performance. We extend the infuser to make several other changes. The \texttt{BRTARGET} and \texttt{MARKLOOP} instructions have already been discussed, and the \texttt{INVOKELIGHT} instruction is the topic of the next section. In addition to these, we made the following four other modifications to Darjeeling's instruction set:

\subsubsection{\texttt{GET/PUTFIELD\_A\_FIXED} reference field access}
\begin{figure}[]
  \makebox[\hsize][c]{\includegraphics[width=0.8\linewidth]{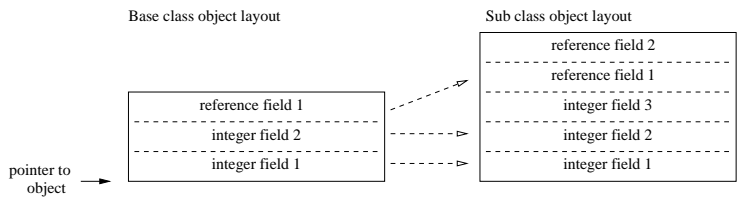}}
  \caption{Base class and sub class layout }
  \label{fig-super-class-sub-class-field-layout}
\end{figure}

The \texttt{GETFIELD\_*} and \texttt{PUTFIELD\_*} instructions are used to access fields in objects. Because of Darjeeling's split architecture, the offset from the object pointer is known at compile time only for integer fields, but not for reference fields. As shown in Figure \ref{fig-super-class-sub-class-field-layout}, integer fields will be at the same offset, regardless of whether an object is of the compile-time type, or a subclass. References fields may shift up in subclass instances, so \texttt{GETFIELD\_A} and \texttt{PUTFIELD\_A} must examine the object's actual type and calculate the offset accordingly, adding significant overhead.

This overhead can be avoided if we can be sure of the offset at compile time, which is the case if the class is marked \texttt{final}. In this case the infuser will replace the \texttt{GETFIELD\_A} or \texttt{PUTFIELD\_A} opcode with a \texttt{\_FIXED} version so the VM knows it is safe to determine the offset at AOT compile time. Conveniently, one of the optimisations ProGuard does, is marking any class that is not subclassed as \texttt{final}, so most of this is automatic.

\paragraph{Alternative solutions} An alternative we considered is to let go of Darjeeling's split architecture for object fields and mix them, so the offsets for reference fields would also be known at compile time. To allow the garbage collector to find the reference fields we could either extend the class descriptors with a bit map indicating the type of each slot, or let the garbage collector scan all classes in the inheritance line of an object.

We chose our solution because it is easy to implement and adds only a few bytes to the VM size, while the garbage collector is already one of the most complex components of the VM. Also, we found that almost all classes in our benchmark could be marked \texttt{final}. But either solution would work, and the alternative could be considered as a more general solution.

\paragraph{Evaluation}
The impact of this optimisation is significant, but we decided not to include it in our evaluation since the overhead is the result of implementation choices in Darjeeling, which was optimised for size rather than performance. This means the overhead is rather arbitrary, and not a direct result of the AOT techniques or the JVM's design. Therefore, all results reported in this paper are with this optimisation already turned on.

Since Darjeeling's split architecture has a lot of advantages in terms of complexity and VM size, we still feel it is important to mention this as an example of the kind of trade-offs faced when optimising for performance.

\subsubsection{\texttt{SIMUL} 16-bitx16-bit to 32-bit multiplication}
While Darjeeling already introduced 16-bit arithmetic operations, it does not cover the case of multiplying two 16-bit shorts, and storing the result in a 32-bit integer. In this case the infuser would emit \texttt{S2I} instructions to convert the operands to two 32-bit integers, and then use the normal \texttt{IMUL} instruction for full 32-bit multiplication. On a device with a shorter word size, this is significantly more expensive than 16x16 to 32-bit multiplication.

We added a new opcode, \texttt{SIMUL}, for this case, which the infuser will emit if it can determine the operands are 16-bit, but the result is used as a 32-bit integer.

We could added more instructions, for example \texttt{SIADD} instruction for addition, \texttt{BSMUL} for 8-bit to 16-bit multiplication, etc. But there is always a trade-off between the added complexity of an optimisation and the performance improvement it yields, and for these cases this is much smaller than for \texttt{SIMUL}.

\subsubsection{16-bit array indexes}
Normal JVM array access instructions (\texttt{IASTORE}, \texttt{IALOAD}, etc) expect the index operand to be a 32-bit integer. On a sensor node with only a few KB of memory, we will never have arrays that require such large indexes, so we modified the array access instructions to expect a 16-bit index instead. This is easily done in Darjeeling's infuser, which contains a specification of the type of operands of each opcode, and will automatically emit type conversions where necessary.

This complements one of the manual optimisations discussed in Section \ref{sec-optimisations-manual-java-source-optimisation}. Using short values as index variables makes operations on the index variable cheaper, while changing the operand of the array access instructions reduces the amount of work the array access instruction needs to do and the number of registers it requires.

\subsubsection{Constant bit shifts}
\label{sec-opt-constant-shift}
Finally, shifts by a constant number of bits appear in seven of the eight benchmarks described in Section 6. They appear not only in computation intensive benchmarks, but also as optimised multiplications or divisions by a power of 2, which are common in many programmes.

In JVM bytecode the shift operators take two operands from the stack: the value to shift, and the number of bits to shift by. While this is generic, it is not efficient for constant shifts: we first need to push the constant onto the stack, and then the bit shift is implemented as a simple loop which shifts one bit at a time. If we already know the number of bits to shift by, we can generate much more efficient code.

Note that this is different from other arithmetic operations with a constant operand. For operations such as addition, our translation process results in loading the constant and performing the operation, similar to what \texttt{avr-gcc} produces in most cases. An addition takes just as long when the operand is taken from the stack, as when it is a constant. What makes bit shifts a special case is that for an unknown number of bits a loop must be generated to shift one bit at a time, which is much slower than the code we can generate for a shift by a constant number of bits.

We optimise these cases by adding \texttt{\_CONST} versions of the bit shift instructions \texttt{ISHL}, \texttt{ISHR}, \texttt{IUSHR}, \texttt{SSHL}, \texttt{SSHR}, and \texttt{SUSHR}. We add a simple scan to the infuser to find constant loads that are immediately followed by a bit shift. For these cases the constant load is removed, and the bit shift instruction, for example \texttt{ISHL}, is replaced by \texttt{ISHL\_CONST}, which has a one byte operand containing the number of bits to shift by. On the VM side, implementing these six \texttt{\_CONST} versions of the bit shift opcodes adds 470 bytes to the VM, but it improves performance, sometimes very significantly, for all but one of our benchmarks.

Surprisingly, when we first implemented this, one benchmark performed better than native C. We found that \texttt{avr-gcc} does not optimise constant shifts in all cases. Since our goal is to examine how close a sensor node VM can come to native performance, it would be unfair to include an optimisation that is not found in the native compiler, but could easily be added. We implemented a version that is close to what \texttt{avr-gcc} does, but never better. We only consider cases optimised by \texttt{avr-gcc}. For these, we first emit whole byte moves if the number of bits to shift by is 8 or more, followed by single bit shifts for the remainder. As mentioned before, this optimisation was already included in the example from Table \ref{tbl-simplestackcaching} on, so the effect can be seen by comparing the \texttt{SCONST\_1} and \texttt{SUSHR} instructions in Table \ref{tbl-basic-translation} and the \texttt{SUSHR\_CONST} instruction in Table \ref{tbl-simplestackcaching}.

\section{Optimisations: Method calls}
\label{sec-optimisations-method-calls}

Finally we will look at the overhead caused by method calls. In native code, the smallest functions only need 8 cycles for a \texttt{CALL} and \texttt{RET}, and some \texttt{MOV}s may be needed to move the parameters to the right registers. More complicated functions may spend up to 76 cycles saving and restoring call-saved registers. As we have seen in Section \ref{sec-overhead-method-call}, in Java a considerable amount of state needs to be initialised. For the simplest method call this takes about 550 cycles, and this increases further for large methods with many parameters.

\begin{figure*}[]
  \makebox[\hsize][c]{\includegraphics[width=.5\linewidth]{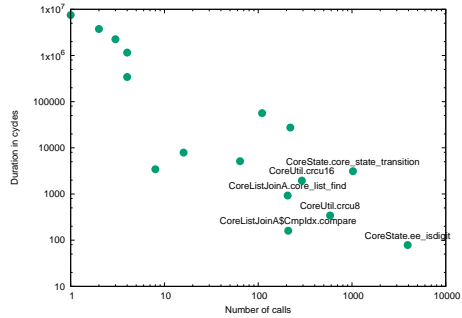}}
  \caption{CoreMark method calls vs duration}
  \label{fig-coremark-method-calls-vs-duration}
\end{figure*}

When we look at the methods in a programme, we typically see a spectrum from a few large methods at the base of the call tree that take a long time to complete and are only called a few times, to small (near-)leaf methods that are fast and frequently called. Figure \ref{fig-coremark-method-calls-vs-duration} shows this spectrum for the CoreMark benchmark.

For the slow methods at the base, the impact of the method call is not very significant for the overall execution time and we can afford to take the 550 cycles penalty. However, as we get closer to the leaf methods, the number of calls increases, as does the impact on the overall performance.

At the very end of this spectrum we have tiny helper functions that may be inlined. However, this is only possible for small methods, or methods called from a single place. In CoreMark's case, \texttt{ee\_isdigit} was small enough to inline. When we inline larger methods, the tradeoff is an increase in code size. So we have a problem in the middle of the spectrum: methods that are too large to inline, but called often enough for the method call overhead to have a significant impact the overall performance.

\subsection{Lightweight methods}
For these cases we introduce a new type of method call: lightweight methods. These methods differ from normal methods in two ways:
\begin{itemize}
	\item we do not create a stack frame for lightweight methods, but use the caller's frame
	\item parameters are passed on the stack, rather than in local variables
\end{itemize}

Lightweight methods give us third choice, in between a normal method call and method inlining. When calling a lightweight method, we directly \texttt{CALL} the method's code. We bypass the VM completely, reusing the caller's stack frame, and leaving the parameters on the (caller's) stack. In effect, the lightweight method behaves similar to inlined code, but since we can \texttt{CALL} it from multiple places, we do not incur the code size overhead of inlining large methods.

Because the method will be called from multiple locations which may have different cache states, we do have to flush the stack cache to memory before a call. This results in slightly more overhead than for inlined code, but much less than for a normal method call.

As an example, consider the simple isOdd method in Listing \ref{lst-lightweight-stack-only}:

\begin{listing}[H]
 \centering
 \begin{minipage}[t]{0.45\textwidth}
  \centering
  \begin{minted}[fontsize=\scriptsize]{java}
// JAVA

public static boolean isOdd(short a) {
    return (a & (short)1)==1;
}
  \end{minted}
 \end{minipage}\hfill
 \begin{minipage}[t]{0.26\textwidth}
  \centering
  \begin{minted}[fontsize=\scriptsize]{java}
// NORMAL METHOD
//           (Stack)
SLOAD_0      (Int)
SCONST_1     (Int,Int)
SAND         (Int)
SRETURN      ()
  \end{minted}
 \end{minipage}\hfill
 \begin{minipage}[t]{0.26\textwidth}
  \centering
  \begin{minted}[fontsize=\scriptsize]{java}
// LIGHTWEIGHT METHOD
//           (Stack)
SCONST_1     (Int,Int)
SAND         (Int)
SRETURN      ()
  \end{minted}
 \end{minipage}
\caption{Simple, stack-only lightweight method example}
\label{lst-lightweight-stack-only}
\end{listing}

The normal implementation has a single local variable. It expects the parameter to be stored there and the stack to be empty when we enter the method. In contrast, the lightweight method does not have any local variables and expects the parameter to be on the stack.

We added a new instruction, \texttt{INVOKELIGHT}, to call lightweight methods. In the bottom half of Listing \ref{lst-comparison-lightweight-and-normal-invocation} we see how \texttt{INVOKELIGHT} and \texttt{INVOKESTATIC} are translated to native code. Both first flush the stack cache to memory. After that, the lightweight method can directly call the implementation of isOdd, while the native version first saves the stack pointers, and then enters an expensive call into the VM to setup a stack frame for \texttt{isOdd}, which in turn will call the actual method.

\begin{listing}[H]
 \centering
 \begin{minipage}[t]{0.39\textwidth}
  \centering
  \begin{minted}[fontsize=\scriptsize]{java}
// NORMAL INVOCATION
// INVOKESTATIC isOdd:
  push r25        // Flush the cache
  push r24
  call &preinvoke // Save X and SP
  ldi r22, 253    // Set parameters
  ldi r23, 2      //  for callMethod
  ldi r24, 21
  ldi r20, 64
  ldi r21, 42
  ldi r18, 13
  ldi r19, 0
  ldi r25, 2
  call &callMethod // Call to VM
  call &postinvoke // Restore X and SP
  \end{minted}
 \end{minipage}
 \begin{minipage}[t]{0.29\textwidth}
  \centering
  \begin{minted}[fontsize=\scriptsize]{java}
// LIGHTWEIGHT INVOCATION
// INVOKELIGHT isOdd:
  push r25      // Flush the cache
  push r24
  call &isOdd
  \end{minted}
 \end{minipage}
\caption{Comparison of lightweight and normal method invocation}
\label{lst-comparison-lightweight-and-normal-invocation}
\end{listing}

\subsubsection{Local variables}
The lightweight implementation of the \texttt{isOdd} example only needs to process the values that are on the stack, but this is only possible for the smallest methods. If we want a lightweight method to be able to use local variables, we need to reserve space for them in the caller's stack frame, equal to the maximum number of slots needed by all the lightweight methods it may call.

In our AOT compiled code, we use the ATmega's Y register to point the start of a method's local variables. To call a lightweight method with local variables, the caller only needs to shift Y up to the region reserved for lightweight method variables before doing the \texttt{CALL}. The lightweight method can then access its locals as if it were a normal method.

\subsubsection{Nested calls}
A final extension is to allow for nested calls. While frequently called leaf methods benefit the most from lightweight methods, there are many cases where it is useful for lightweight methods to call other lightweight methods. A good example from the CoreMark benchmark is the 16-bit \texttt{crcu16} function, which is implemented as two calls to \texttt{crcu8}. While \texttt{crcu8} is the most critical, there is still one call to \texttt{crcu16} for every two to \texttt{crcu8}.

So far we have not discussed how to handle the return address in a lightweight method. Our AOT compiler uses the native stack to store JVM integer stack value, which means the operands to a lightweight method will be on the native stack. But when we do a \texttt{CALL}, the return address is put on the stack, covering the method parameters.

For leaf methods, the lightweight method will first pop the return address into two fixed registers, and avoid using these register for stack caching. When the method returns, the return address is pushed back onto the stack before the \texttt{RET} instruction.

For lightweight methods that will call another lightweight method, the return value is also popped from the stack, but instead of leaving it in the fixed register, where it would be overwritten by the nested call, we save it in the first local variable slot and increment Y to skip this slot. Since each lightweight method has its own block of locals, we can nest calls as deeply as we want.

This difference in method prologue and epilogue is the only difference in the way the VM generates code for a lightweight method, all JVM instructions can then be translated the same way as for a normal method.

\subsubsection{Stack frame layout}
A normal method that invokes a possible string of lightweight methods, needs to save space for this in its stack frame. How much space it needs to reserve can be determined by the infuser at compile time, and this information is added to the method descriptor.

\begin{figure}[]
  \makebox[\hsize][c]{\includegraphics[width=0.80\linewidth]{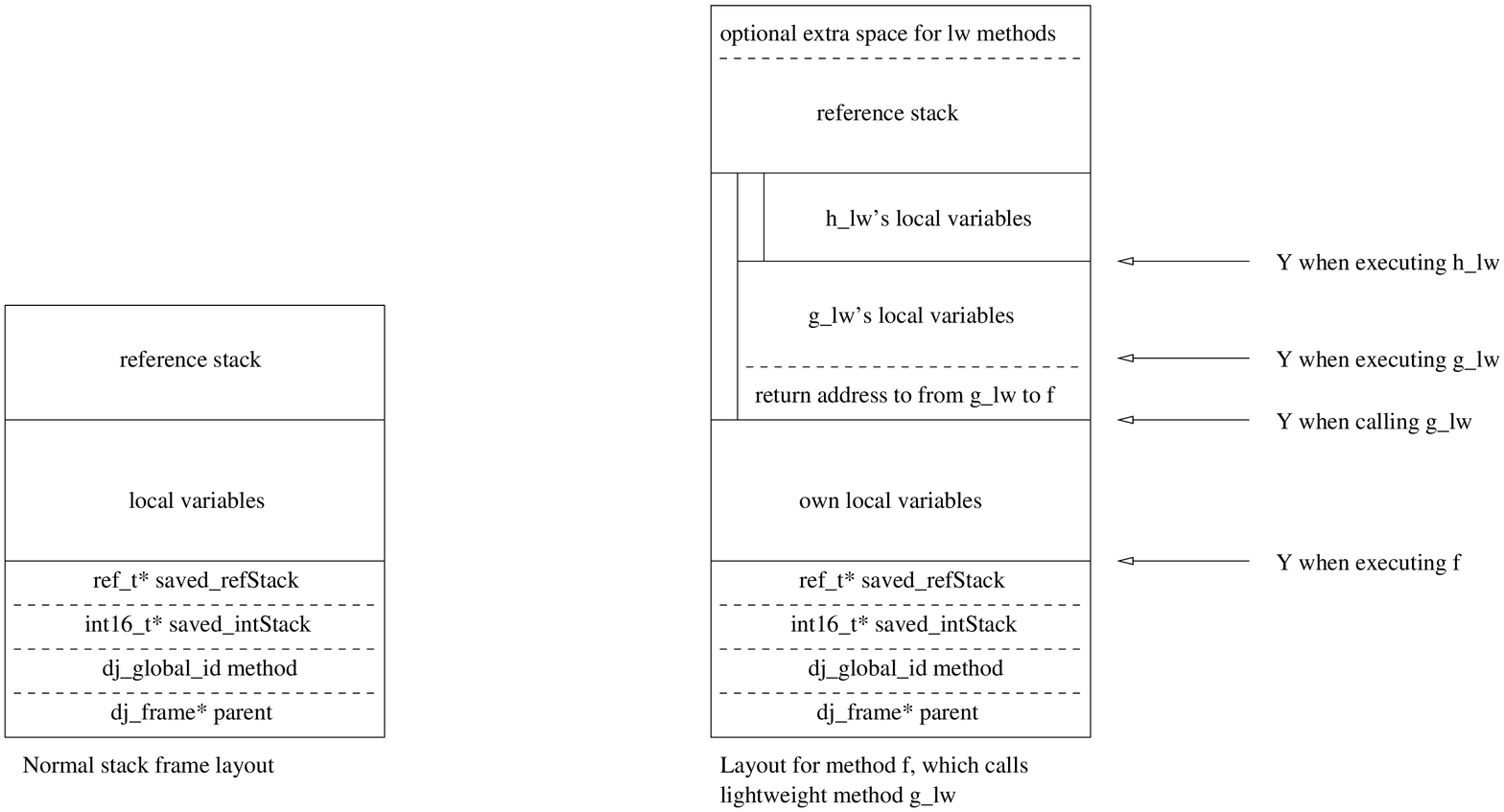}}
  \caption{Stack frame layout for a normal method \texttt{f}, which calls lightweight method \texttt{g\_lw}, which in turn calls lightweight method \texttt{h\_lw}.}
  \label{fig-stack-frame-lightweight-method}
\end{figure}

An example is shown in Figure \ref{fig-stack-frame-lightweight-method}, which shows the stack frame for a normal method \texttt{f}, which calls lightweight method \texttt{g\_lw}, which in turn calls another lightweight method \texttt{h\_lw}.

The stack frame for \texttt{f} contains space for its own locals, and for the locals of the lightweight method it calls: \texttt{g\_lw}. In turn, \texttt{g\_lw}'s locals contain space for \texttt{h\_lw}'s locals, as well as a slot to store the return address back to \texttt{f}. Since \texttt{h\_lw} does not call any other methods, it just keeps its return address in registers.

When a method calls a lightweight method with local variables, it will move the Y register to point at that method's locals. From Figure \ref{fig-stack-frame-lightweight-method} it is clear it only needs to increment Y by the size of its own locals. For \texttt{f}, this will place the Y register at the beginning of \texttt{g\_lw}'s locals. Since \texttt{g\_lw} may call \texttt{h\_lw}, \texttt{g\_lw}'s prologue will first store its return address in the first local slot, moving Y forward in the process so that Y points to the first free slot.

\subsubsection{Mark loop}
Lightweight methods may use any register and do not save call-saved registers like normal methods. The only case where this would be necessary is when it is called inside a \texttt{MARKLOOP} block that uses the same register to pin a variable. In this case we save those variables back to memory before calling the lightweight method and load them after the call returns. Since lightweight methods always come before their invocation in the infusion, the VM already knows which registers it uses, and will only save and restore pinned variables if there is a conflict. Because registers for mark loop are allocated low to high, and for normal stack caching from high to low, in many cases the two may not collide.

\subsubsection{Example call}
An example of the most complex case for a lightweight call is shown in Listing \ref{lst-full-lighweight-method-call}, which shows how method \texttt{f} from Figure \ref{fig-stack-frame-lightweight-method} would call \texttt{g\_lw}, assuming \texttt{f} is in a markloop block at the time which pinned a variable R14:R15, and these registers are also used by \texttt{g\_lw}.

In the translation of the \texttt{INVOKELIGHT} instruction we see we first flush the cache to memory, and then save the value of the local short at offset 22 that was pinned to R14:R15. Finally we add 26 to the Y register to skip the caller's own local variables and point Y to the start of the space reserved for lightweight method locals.

In the method call, we first see the return address is popped off the stack into a register. Since \texttt{g\_lw} may call another lightweight method, we cannot leave it there but store it in the first local slot, incrementing Y in the process. After \texttt{g\_lw} is done, we see the reverse process to return to the caller, where we then see the Y register is restored to point to the caller's locals, and the local variable at offset 22 is loaded back into the pinned register.

\begin{listing}[H]
 \centering
 \begin{minipage}[t]{0.45\textwidth}
  \begin{minted}[fontsize=\scriptsize]{java}
// LIGHTWEIGHT INVOCATION
INVOKELIGHT g_lw
 push r25        // Flush the cache
 push r24
 std Y+22, r14   // Save pinned value
 std Y+23, r15
 adiw Y, 26      // Move Y to g_lw's locals
 call &g_lw
 sbiw Y, 26      // Restore Y
 ldd r14, Y+22   // Reload pinned value
 ldd r15, Y+23
  \end{minted}
 \end{minipage}\hfill
 \begin{minipage}[t]{0.45\textwidth}
  \centering
  \begin{minted}[fontsize=\scriptsize]{java}
// IMPLEMENTATION OF g_lw
pop r18     // Pop the return address
pop r19
st Y+, r18  // Save in 1st local,
st Y+, r19  //  and increment Y

  .. // g_lw's body

ld r19, -Y  // Load return address,
ld r18, -Y  //  and decrement Y
push r19    // Push return address
push r18    //  onto the stack
ret
  \end{minted}
 \end{minipage}
\caption{Full lightweight method call}
\label{lst-full-lighweight-method-call}
\end{listing}

\subsection{Overhead comparison}
We now compare the overhead for the various ways we can call a method in Table \ref{tbl-method-invoke-overhead-comparison}.

Manually inlining code yields the best performance, but at the cost of increasing code size if larger methods are inlined. ProGuard inlining is currently slightly expensive because of the way it always saves parameters in local variables.

Both lightweight methods options cause some overhead, although this is very little compared to a full method call. First, we need to flush the stack cache to memory to make sure the parameters are on the real stack. This this takes two \texttt{push} and eventually two corresponding \texttt{pop} instructions per word, costing 8 cycles per word. In addition, we need to clear the value tags from the stack cache, which may mean we may not be able to skip as many \texttt{LOAD} instructions after the lightweight call, but this is hard to quantify.

Next the cost of translating the \texttt{INVOKE} instruction varies depending on the situation. In the simplest case it is simply a \texttt{CALL} to the lightweight method, which together with the corresponding \texttt{RET} costs 8 cycles. The worst case is 68 cycles when the lightweight method has local variables, uses all registers, and the caller used the maximum of 7 pairs to pin variables in a \texttt{MARKLOOP} block.

After calling the method, the method prologue for lightweight methods is very simple. We just need to save the return address and restore it in the epilogue, which takes 8 cycles if we can leave it in a register, or 16 if we need to store it in a local variable slot.

For small handwritten lightweight methods this is the only cost, but for larger ones created by converting a Java method, we add \texttt{STORE} instructions to copy the parameters from the stack into local variables, as shown in Listing \ref{lst-comparison-of-handwritten-and-converted-java-lw-method}. This is similar to the only overhead incurred by ProGuard's method inlining, and costs 4 cycles per word for the \texttt{STORE}, and possibly 4 more if the corresponding \texttt{LOAD} cannot be eliminated by popped value caching.

The total overhead for a lightweight method call scales nicely with the method's complexity. For the smallest methods, the minimum is only 16 cycles, plus 8 cycles per word for the parameters. For the most complex cases this may go up to 100 to 150 cycles. But these methods must be more complex and will have a longer run-time, so the relative overhead is still acceptable.

The number of cycles in Table \ref{tbl-method-invoke-overhead-comparison} is just a broad indication of the overhead. Some factors, such as the cost of clearing the value tags is hard to predict, and inlining may allow some optimisations that aren't possible with a method call. In practice the actual cost in a number of specific cases we examined varies, but is in the range we predicted.

Comparing this to a normal method call, we see the cost is much higher, and less dependent on the complexity of the method that is called. The overhead from setting up the stack frame, and the more expensive translation of the \texttt{INVOKE} instruction (see Listing \ref{lst-lightweight-stack-only}) are fixed, meaning a call will cost at least around 550 cycles, increasing to over 700 cycles for more complex methods taking many parameters.

\begin{table*}[]
\begin{minipage}{\textwidth} 
\centering
\caption{Approximate cycles of overhead caused by different ways of invoking a method}
\label{tbl-method-invoke-overhead-comparison}
\small
\makebox[\hsize][c]{\begin{tabular}{lccccc}
\toprule
                                                          & Manual          & ProGuard                & Stack-only                  & Converted Java              & Normal                                              \\
                                                          & inlining        & inlining                & lightweight                 & lightweight                 & method call                                         \\
\hline
flush the stack cache \footnote{excluding effect on future popped value cache performance because of cleared value tags}
                                                          &                 &                         & 8 per word                  & 8 per word                  & 8 per word                                          \\
\texttt{INVOKE}                                           &                 &                         & 8 to 68                     & 8 to 68                     & \textasciitilde 82                                  \\
create stack frame                                        &                 &                         &                             &                             & \textasciitilde 450                                 \\
method pro-/epilogue                                      &                 &                         & 8 or 16                     & 8 or 16                     & 10 to 71                                            \\
store and load parameters                                 &                 & 4 or 8 per word         &                             & 4 or 8 per word             & 4 or 8 per word                                     \\
\\
\emph{total}                                              &                 & \emph{4 or 8 per word}  & \emph{16 to 84 +}           & \emph{16 to 84 +}           & \emph{\textasciitilde 542 to \textasciitilde 603 +} \\
                                                          &                 &                         & \emph{8 per word}           & \emph{12 or 16 per word}    & \emph{12 or 16 per word}                            \\
\bottomrule
\end{tabular}}
\end{minipage}
\end{table*}

\subsection{Creating lightweight methods}
We currently support two ways to create a lightweight method:
\begin{itemize}
	\item handwritten JVM bytecode
	\item converting a Java method 
\end{itemize}

\subsubsection{Handwritten JVM bytecode}
For the first option we declare the methods \texttt{native} in the Java source code, so the code calling it will compile as usual. We provide the infuser with a handwritten implementation in JVM bytecode, which the infuser will simply add to the infusion, and then process it in the same way as it processes a normal method, with one step added:

For lightweight methods, the parameters will be on the stack at the start of the method, but the infusers expects to start with an empty stack. To allow the infuser to process them like other methods, we add a dummy \texttt{LW\_PARAMETER} instruction for each parameter. This instruction is skipped when writing the binary infusion, but it tricks the infuser into thinking the parameters are being put on the stack.

\subsubsection{Converting Java methods}
This handwritten approach is useful for the smallest methods, and allows us to create bytecode that only uses the stack, which produces the most efficient code. But for more complex methods it quickly becomes very cumbersome to write the bytecode by hand.

As a second, slightly slower, but more convenient option, we developed a way to convert normal Java methods to lightweight methods by adding a \texttt{@Lightweight} annotation to it.

The infuser will scan all the methods in an infusion for this annotation. When it finds a method marked \texttt{@Lightweight}, the transformation to turn a normal JVM method into a lightweight one is simple: we first add a dummy \texttt{LW\_PARAMETER} instruction for each parameter, followed by \texttt{STORE} instructions to pop these parameters off the stack and store them in the right local variables. After this, we can use the normal body of the method and call it as a lightweight method.

Listing \ref{lst-comparison-of-handwritten-and-converted-java-lw-method} shows the difference for the \texttt{isOdd} method. We can see this approach adds some overhead in the form of a \texttt{SSTORE\_0} and a \texttt{SLOAD\_0} instruction. However, using popped value caching, only the \texttt{SSTORE\_0} will have a run-time cost. Another disadvantage of the converted method is that is has to use a local variable, which will slightly increase memory usage, but in return this approach gives us a very easy way to create lightweight methods.

\begin{listing}[H]
 \centering
 \begin{minipage}[t]{0.24\textwidth}
  \centering
  \begin{minted}[fontsize=\scriptsize]{java}
// HANDWRITTEN
//            (Stack)
LW_PARAMETER  (Int)
SCONST_1      (Int,Int)
SAND          (Int)
SRETURN       ()
  \end{minted}
 \end{minipage}\hfill
 \begin{minipage}[t]{0.40\textwidth}
  \centering
  \begin{minted}[fontsize=\scriptsize]{java}
// JAVA
@Lightweight
public static boolean isOdd(short a) {
    return (a & (short)1)==1;
}
  \end{minted}
 \end{minipage}\hfill
 \begin{minipage}[t]{0.25\textwidth}
  \centering
  \begin{minted}[fontsize=\scriptsize]{java}
// CONVERTED JAVA
//            (Stack)
LW_PARAMETER  (Int)
SSTORE_0      ()
SLOAD_0       (Int)
SCONST_1      (Int,Int)
SAND          (Int)
SRETURN       ()
  \end{minted}
 \end{minipage}
 \vspace{0.5cm}

\caption{Comparison of hand written lightweight method and converted Java method}
\label{lst-comparison-of-handwritten-and-converted-java-lw-method}
\end{listing}

\subsubsection{Replacing \texttt{INVOKE}s}
The infuser does a few more transformations to the bytecode. Every method is scanned for \texttt{INVOKESTATIC} instructions that invoke a lightweight method. These are simply replaced by an \texttt{INVOKELIGHT}, and the number of extra slots for the reference stack and local variables of the current method is increased if necessary. Finally, methods are sorted so a lightweight method will be defined before it is invoked, to make sure the VM can always generate the \texttt{CALL} directly.

\subsection{Limitations and tradeoffs}
There are a few limitations to the use of lightweight methods:
\paragraph{No recursion} Since we need to be able to determine how much space to reserve in the caller's stack frame for a lightweight method's reference stack and local variables, we do not support recursion, although lightweight calls can be nested.

\paragraph{No garbage collection}
Lightweight methods reuse the caller's stack frame. This is a problem for the garbage collector, which works by inspecting each stack frame and finding the references on the stack and in local variables. If the garbage collector would be triggered while we're in a lightweight call, it would not know where to find the lightweight method's references, since the stack frame only has information for the method that owns it.

While it may be possible to relax this constraint with some effort, in most cases this is only a minor restriction. Lightweight methods are most useful for fast and frequently called methods, and operations that may trigger the garbage collector are usually expensive, so there is less to be gained from using a lightweight method in these situations.

\paragraph{Static only}
We currently do not support lightweight virtual methods, since the overhead of resolving the target of the invoke is large compared to the rest of the invoke overhead, but this is something that could be considered in future work.

\paragraph{Stack frame usage}
Finally, while many methods can be made lightweight, we should remember that a method calling a lightweight method will always reserve space for it in its locals. This space is reserved, regardless of whether the method is currently executing or not, and the more nested lightweight calls are made, the more space we need to reserve.

As an example if we have a method \texttt{f1} which may call a lightweight method with a large number of local variables, \texttt{big\_lw}, but is currently calling normal method \texttt{f2}, which may also call \texttt{big\_lw}, we will have reserved space for \texttt{big\_lw} twice, both in \texttt{f1}'s and in \texttt{f2}'s frame.

\section{Evaluation}
We use a set of eight different benchmarks to measure the effect of our optimisations:

\begin{itemize}
\item \emph{bubble sort}: taken from the Darjeeling sources, and used in \cite{Brouwers:2009cj, Ellul:2012thesis}
\item \emph{heap sort}: standard heap sort \cite{heapsort}
\item \emph{binary search}: taken from the TakaTuka \cite{Aslam:2008} source code
\item \emph{fft}: fixed point FFT, adapted from the widespread fix\_fft.c
\item \emph{xxtea}: as published in \cite{Wheeler:1998}
\item \emph{md5}: also taken from the Darjeeling sources, and used in \cite{Brouwers:2009cj, Ellul:2012thesis}
\item \emph{rc5}: from LibTomCrypt \cite{libtomcrypt}
\item \emph{CoreMark 1.0}: a freely available benchmark developed by EEMBC \cite{coremark}
\end{itemize}

The first seven are small benchmarks, consisting of only one or two methods. They all process an array of data, which we expect to be common on a sensor node, and likely to be a performance sensitive operation. However, the processing they do is different for each benchmark, allowing us to examine how our optimisations respond to different kinds of code. The eighth benchmark, CoreMark, is a standard benchmark representative of larger embedded applications.

For each benchmark we implemented both a C and a Java version, keeping both implementations as close as possible. We manually optimised the code as described in Section \ref{sec-optimisations-manual-java-source-optimisation}. These optimisations did not affect the performance of the C version, indicating \texttt{avr-gcc} already does similar transformations on the original code. We use \texttt{javac} version 1.8.0, ProGuard 5.2.1, and \texttt{avr-gcc} version 4.9.1. The C benchmarks are compiled at optimisation level -O3, the rest of the VM at -Os.

We manually examined the compiled code produced by \texttt{avr-gcc}. While we identified some points where more efficient code could have been generated, except for the constant shifts mentioned in the previous section, this did not affect performance by more than a few percent. This leads us to believe \texttt{avr-gcc} is a fair benchmark to compare to.

We run our VM in the cycle-accurate Avrora simulator \cite{Titzer:2005vb}, emulating an ATmega128 processor. We modified Avrora to get detailed traces of the compilation process and of the run-time performance of both C and AOT compiled code.

Our main measurement for both code size and performance is the overhead compared to optimised native C. To compare different benchmarks, we normalise this overhead to a percentage of the number of bytes or cpu cycles used by the native implementation: a 100\% overhead means the AOT compiled version takes twice as long to run, or twice as many bytes to store. The exact results can vary depending on factors such as which benchmarks are chosen, the input data, etc., but the general trends are all quite stable.

\subsection{CoreMark}
\label{sec-evaluation-coremark}

First, we will examine the CoreMark benchmark. CoreMark was developed by the Embedded Microprocessor Benchmark Consortium as a general benchmark for embedded CPUs. It consists of three main parts:
\begin{itemize}
  \item matrix multiplication
  \item a state machine
  \item linked list processing
\end{itemize}

As mentioned before, we kept the Java versions as close to the original C code as possible. The other benchmarks are all relatively simple, and porting them to Java is straightforward. CoreMark is a much more comprehensive benchmark, and the more complex code exposes some challenges when using Java on embedded devices.

The biggest complication is that CoreMark makes extensive use of pointers, which do not exist in Java. In cases where a pointer to a simple variable is passed to a function, we simply wrap it in a wrapper object. A more complicated case is the \texttt{core\_list\_mergesort} function, which takes a function pointer parameter \texttt{cmp} used to compare list elements. Two different implementations exists, \texttt{cmp\_idx} and \texttt{cmp\_complex}. Here we choose the most canonical way to do this in Java, which is to define an interface and pass an object with the right to implementation \texttt{core\_list\_mergesort}.

Finally, the C version of the linked list benchmark takes a block of memory and constructs a linked list inside it by and treating it as \texttt{list\_head} and \texttt{list\_data} structs, shown in Listing \ref{lst-coremark-list-data-structures}. One way to mimic this as closely as possible is to use an array of shorts of equal size to the memory block used in the C version, and use indexes into this array instead of pointers. However this leads to quite messy code.

Instead we choose the more natural Java approach and define two classes to match the structs in C and create instances of these to initialise the list. This is also the faster option because accessing fields is faster than array access. The trade-off is memory consumption, since each object has its own heap header.

\begin{listing}[H]
 \centering
 \begin{minipage}[t]{0.45\textwidth}
  \centering
  \begin{minted}[fontsize=\scriptsize]{c}
typedef struct list_data_s {
    ee_s16 data16;
    ee_s16 idx;
} list_data;

typedef struct list_head_s {
    struct list_head_s *next;
    struct list_data_s *info;
} list_head;
  \end{minted}
 \end{minipage}\hfill
 \begin{minipage}[t]{0.45\textwidth}
  \centering
  \begin{minted}[fontsize=\scriptsize]{java}
public static final class ListData {
    public short data16;
    public short idx;
    }

public static final class ListHead {
    ListHead next;
    ListData info;
}
  \end{minted}
 \end{minipage}
\caption{C and Java version of the CoreMark list data structures}
\label{lst-coremark-list-data-structures}
\end{listing}

\subsubsection{Manual optimisations}
After translating the C to Java code, we only do 'fair' manual optimisations that we believe a future optimising infuser could easily do automatically. Since CoreMark is our most comprehensive benchmark, we use it to evaluate the effect of these manual optimisations.

\begin{table*}[]
\begin{minipage}{\textwidth} 
\centering
\caption{Effect of manual source optimisation on the CoreMark benchmark}
\label{tbl-coremark-manual-optimisation}
\small
\makebox[\hsize][c]{\begin{tabular}{lrrrrrrrr}
\toprule
                                                                     & \multicolumn{2}{l}{list}     & \multicolumn{2}{l}{matrix}     &  \multicolumn{2}{l}{state}       & \multicolumn{2}{l}{total}     \\
                                                                     & \tiny time \footnote{in millions of cycles} & \tiny vs nat. C & \tiny time  & \tiny vs nat. C& \tiny time  & \tiny vs nat. C& \tiny time  & \tiny vs nat. C \\
\hline
native C                                                             &        17.9 &                &          49.8 &                &        18.4   &                  &         86.0 &                \\
baseline                                                             &       122.0 &       (+583\%) &         367.8 &       (+639\%) &       293.7   &       (+1496\%)  &        783.5 &       (+811\%) \\
optimised, using original source                                     &        52.6 &       (+195\%) &         239.1 &       (+380\%) &        82.8   &        (+350\%)  &        374.4 &       (+335\%) \\
\makebox[5mm]{} \tiny manually inline small methods                  & \tiny  -0.4 & \tiny   (-3\%) & \tiny   -37.2 & \tiny  (-75\%) & \tiny -17.0   & \tiny   (-92\%)  & \tiny  -54.5 & \tiny  (-63\%) \\
\makebox[5mm]{} \tiny use short array index variables                & \tiny  +6.8 & \tiny  (+39\%) & \tiny  -109.4 & \tiny (-219\%) & \tiny  -5.0   & \tiny   (-27\%)  & \tiny -107.6 & \tiny (-125\%) \\
\makebox[5mm]{} \tiny avoid recalculating expressions   in  a loop   & \tiny   0.0 & \tiny   (-1\%) & \tiny    -7.6 & \tiny  (-15\%) & \tiny   0.0   & \tiny    ( 0\%)  & \tiny   -7.6 & \tiny   (-9\%) \\
\makebox[5mm]{} \tiny reduce array and object access                 & \tiny  -0.2 & \tiny   (-1\%) & \tiny   -18.1 & \tiny  (-37\%) & \tiny  -2.4   & \tiny   (-14\%)  & \tiny  -20.7 & \tiny  (-24\%) \\
\makebox[5mm]{} \tiny reduce branch cost in crcu8                    & \tiny  -4.0 & \tiny  (-22\%) & \tiny    -0.5 & \tiny   (-1\%) & \tiny  -3.5   & \tiny   (-19\%)  & \tiny   -8.1 & \tiny  (-10\%) \\
using optimised source                                               &        54.8 &       (+207\%) &          66.3 &        (+33\%) &        54.9   &        (+198\%)  &        175.9 &       (+104\%) \\
\hline
\makebox[5mm]{} \tiny (unfair) avoid creating objects                & \tiny  -3.4 & \tiny  (-19\%) & \tiny     0.0 & \tiny    (0\%) & \tiny -11.4   & \tiny   (-61\%)  & \tiny  -14.7 & \tiny  (-17\%) \\
\makebox[5mm]{} \tiny (unfair) avoid virtual calls                   & \tiny -22.8 & \tiny (-128\%) & \tiny     0.0 & \tiny    (0\%) & \tiny   0.0   & \tiny     (0\%)  & \tiny  -22.8 & \tiny  (-26\%) \\
after 'unfair' optimisations                                         &        28.6 &        (+60\%) &          66.3 &        (+33\%) &        43.5   &        (+137\%)  &        138.4 &        (+61\%) \\
\bottomrule
\end{tabular}}
\end{minipage}
\end{table*}

Table \ref{tbl-coremark-manual-optimisation} shows the slowdown over the native C version, broken down into CoreMark's three main components. The baseline version, using the original Java code and without using any of our optimisations, is 811\% slower than native C. Even after applying all our other optimisations, the best we can achieve with the original code is a 335\% slowdown, proving the importance of a better optimising infuser.

Next we apply our manual optimisations to the Java source code, as described in Section \ref{sec-optimisations-manual-java-source-optimisation}, and add a small extra optimisation to \texttt{crcu8} which can be easily reorganised to reduce branch overhead.

The effect depends greatly on the characteristics of the code. The matrix part of the benchmark benefits most from using short array indexes, the state machine frequently calls a small method and benefits greatly from inlining it, etc. The reason the linked list part is slightly slower after using short array index variables is that it allocates a small object, and the change in memory usage means this now triggers a run of the garbage collector, which presumably had already happened earlier in the version with int index variables. Combined these optimisations reduce the overhead for the whole benchmark from 335\% to 104\%.

We also applied all these optimisations to the native C version to ensure a fair comparison, but the difference in performance was negligible.

In the rest of the evaluation, all the results presented are for the manually optimised Java code.

\subsubsection{'Unfair' optimisations}
\label{sec-evaluation-coremark-unfair-optimisations}
After these optimisations, CoreMark is still the slowest of our benchmarks, and the only one to still be at more than 100\% overhead. We can improve performance further if we relax our constraint of only doing optimisations that a compiler could do automatically without changing the code significantly.

In Table \ref{tbl-coremark-manual-optimisation} we see that in the native version, over half of the time is spent in the matrix part of the benchmark, but for the final Java version we see all three parts taking roughly the same time. The state machine and linked list processing both suffer from a much larger slowdown than the matrix part, which by itself would be the second fastest of all our benchmarks.

One of the reasons for the slow performance of the state machine is that it creates two arrays of 8 ints, and an little wrapper object for a short to mimic a C pointer. Allocating memory on the Java heap is much more expensive than it is for a local C variable.

The linked list benchmark also creates a small object, but here the biggest source of overhead is in the virtual method call to the compare objects in \texttt{core\_list\_mergesort} that we use instead of a function pointer. Virtual methods cannot be made lightweight.

This is the best we can do when we strictly translate the C to Java code, and only do optimisations that could be done automatically. If we relax this constraint, we can remove these two sources of overhead as well: because we know we the code will not run multithreaded or recurse, we could choose to statically allocate the small objects used by the state machine, and one by the linked list part, since they only use 90 bytes. The virtual call to the comparer objects in the list benchmark is the most natural implementation this in Java, but given that we know there are only two implementations, we can make both compare methods \texttt{static} and pass a boolean to select which one to call instead of the comparer object. This saves the virtual method call, and allows ProGuard to inline the methods since they are only called from a single location.

Combined this improves the performance of CoreMark to only 61\% overhead over native C, right in the middle of the spectrum of the other benchmarks. However, the code is now fundamentally different than the original CoreMark, so it is not a completely fair comparison, although a developer writing this code in Java from the start may have made similar choices.

Either way, these results point at some weaknesses of Java when used as an embedded VM. The lack of cheap function pointers, or a way of allocating small local objects or arrays in a method's stack frame means there will be a significant overhead in situations where the optimisations we used here cannot be applied.

Neither of these two optimisations were used in the rest of the evaluation.

\subsection{AOT translation overhead}
\begin{table}[]
 \centering
 \caption{Performance data per benchmark}
 \label{tbl-performance-per-benchmark}
 \small
 \scriptsize
 \setlength{\tabcolsep}{4pt}
 \makebox[\hsize][c]{\begin{tabular}{lrrrrrrrrrrrr}
\toprule
BENCHMARK                          & b.sort            &  h.sort           & b.srch            & fft               & xxtea             & md5               & rc5               & coremk            & \makebox[0.2mm]{}   & average           \\
\hline
\multicolumn{10}{l}{EXECUTED JVM INSTRUCTIONS (\%)} \\
\xxt Load/Store                    &              79.8 &              72.1 &              58.8 &              57.8 &              50.9 &              43.7 &              41.1 &            55.5   &     &              57.5 \\
\xxt Constant load                 &               0.2 &               8.1 &               9.8 &              10.8 &              12.5 &              19.1 &              17.6 &            10.1   &     &              11.0 \\
\xxt Processing                    &               8.0 &               7.8 &              13.1 &              22.8 &              32.4 &              28.9 &              36.6 &            14.0   &     &              20.5 \\
  \xxxt   math                     & \xt           8.0 & \xt           5.6 & \xt           9.2 & \xt          12.0 & \xt          10.1 & \xt          12.5 & \xt          10.7 & \xt         8.3   &     & \xt           9.6 \\
  \xxxt   bit shift                & \xt           0.0 & \xt           2.3 & \xt           3.9 & \xt           7.5 & \xt           8.1 & \xt           5.4 & \xt           8.0 & \xt         2.2   &     & \xt           4.7 \\
  \xxxt   bit logic                & \xt           0.0 & \xt           0.0 & \xt           0.0 & \xt           3.2 & \xt          14.2 & \xt          11.0 & \xt          17.9 & \xt         3.6   &     & \xt           6.2 \\
\xxt Branches                      &              12.0 &              11.0 &              17.6 &               4.1 &               4.0 &               5.8 &               2.3 &            16.0   &     &               9.1 \\
\xxt Invoke                        &               0.0 &               0.5 &               0.0 &               0.0 &               0.0 &               0.0 &               0.0 &             0.4   &     &               0.1 \\
\xxt Others                        &               0.0 &               0.5 &               0.7 &               4.5 &               0.2 &               2.5 &               2.4 &             4.0   &     &               1.9 \\
\multicolumn{10}{l}{STACK} \\
\xxt Max. stack (bytes)            &                 8 &                 8 &                 8 &                 8 &                24 &                20 &                14 &              18   &     &              13.5 \\
\xxt Avg. stack (bytes)            &               2.6 &               2.9 &               2.8 &               3.0 &              11.8 &               6.3 &               6.8 &             3.2   &     &               4.9 \\
\hline
\multicolumn{10}{l}{PERFORMANCE OVERHEAD BEFORE OPTIMISATIONS (\%)} \\
\xxt Total                         &             496.7 &             351.4 &             430.8 &             522.6 &             251.5 &             226.5 &             123.4 &           359.0   &     &             345.2 \\
  \xxxt push/pop                   & \xt         183.5 & \xt         139.4 & \xt         192.5 & \xt         220.2 & \xt         168.0 & \xt         105.9 & \xt          61.3 & \xt       128.2   &     & \xt         149.9 \\
  \xxxt load/store                 & \xt         200.1 & \xt         144.3 & \xt         180.9 & \xt         132.7 & \xt          42.5 & \xt          43.9 & \xt          28.5 & \xt        91.9   &     & \xt         108.1 \\
  \xxxt mov(w)                     & \xt          10.4 & \xt           2.9 & \xt          -1.2 & \xt           6.2 & \xt           2.4 & \xt           1.7 & \xt          -1.7 & \xt         4.0   &     & \xt           3.1 \\
  \xxxt other                      & \xt         102.7 & \xt          64.8 & \xt          58.7 & \xt         163.5 & \xt          38.6 & \xt          75.0 & \xt          35.3 & \xt       134.9   &     & \xt          84.2 \\
\multicolumn{10}{l}{PERFORMANCE OVERHEAD REDUCTION PER OPTIMISATION (\%)} \\
\xxt Impr. peephole                &            -162.8 &            -118.8 &            -116.3 &             -99.9 &             -61.8 &             -51.7 &             -23.2 &           -61.0   &     &             -86.9 \\
\xxt Stack caching                 &             -22.9 &             -29.5 &             -76.8 &            -129.8 &             -97.3 &             -54.3 &             -38.6 &           -40.0   &     &             -61.1 \\
\xxt Pop. val. caching             &            -116.6 &             -73.3 &             -29.8 &             -52.4 &              -6.9 &             -12.9 &              -8.8 &           -26.0   &     &             -40.9 \\
\xxt Mark loops                    &             -62.3 &             -28.8 &             -84.4 &             -40.2 &              +5.1 &             -10.9 &              -8.1 &           -40.4   &     &             -33.7 \\
\xxt Const shift                   &               0.0 &              -9.2 &             -22.4 &             -80.4 &             -18.5 &             -43.8 &             -20.2 &           -10.2   &     &             -25.6 \\
\xxt 16-bit array index            &             -37.5 &             -25.3 &             -36.5 &             -22.4 &             -13.8 &              -5.5 &              -4.1 &           -39.3   &     &             -23.1 \\
\xxt SIMUL                         &               0.0 &               0.0 &               0.0 &               0.0 &               0.0 &               0.0 &               0.0 &           -36.7   &     &              -4.6 \\
\multicolumn{10}{l}{PERFORMANCE OVERHEAD AFTER OPTIMISATIONS (\%)} \\
\xxt Total                         &              94.6 &              66.5 &              64.6 &              97.5 &              58.3 &              47.4 &              20.4 &           105.4   &     &             69.3 \\
  \xxxt push/pop                   & \xt           0.0 & \xt          -7.0 & \xt           0.0 & \xt           0.0 & \xt          37.4 & \xt           0.1 & \xt           2.9 & \xt         5.1   &     & \xt          4.8 \\
  \xxxt load/store                 & \xt           9.0 & \xt          33.2 & \xt          28.2 & \xt          22.5 & \xt          -2.3 & \xt          20.3 & \xt           4.3 & \xt        17.6   &     & \xt         16.6 \\
  \xxxt mov(w)                     & \xt          10.4 & \xt           3.9 & \xt          10.8 & \xt           4.6 & \xt           5.6 & \xt           2.2 & \xt           0.5 & \xt        10.0   &     & \xt          6.0 \\
  \xxxt other                      & \xt          75.3 & \xt          36.4 & \xt          25.6 & \xt          70.4 & \xt          17.6 & \xt          24.8 & \xt          12.7 & \xt        72.7   &     & \xt         41.9 \\
\bottomrule
\end{tabular}}

 \setlength{\tabcolsep}{6pt}
\end{table}

Next we will look at the effect of our different optimisations for the baseline AOT translation approach, for all eight of our benchmarks.

The trace data produced by Avrora gives us a detailed view into the run-time performance and the different types of overhead. We count the number of bytes and cycles spent on each native instruction for both the native C and our AOT compiled version, and then group them into 4 categories that roughly match the 3 types of AOT translation overhead discussed in Section \ref{sec-overhead-aot-translation}:
\begin{itemize}
	\item \texttt{PUSH},\texttt{POP}: Matches the type 1 push/pop overhead since native code uses almost no push/pop instructions.
	\item \texttt{LD},\texttt{LDD},\texttt{ST},\texttt{STD}: Matches the type 2 load/store overhead and directly shows the amount of memory traffic.
	\item \texttt{MOV},\texttt{MOVW}: For moves the picture is less clear since the AOT compiler emits them for various reasons. Before we introduce stack caching, it emits moves to replace push/pop pairs, and after the mark loops to save a pinned value when it is popped destructively.
	\item others: the total overhead, minus the previous three categories. This roughly matches the type 3 overhead.
\end{itemize}

We define the overhead from each category as the number of bytes or cycles spent in the AOT version, minus the number spent by the native version for that category, and again normalise this to the \emph{total} number of bytes or cycles spent in the native C version. The detailed results for each benchmark and type of overhead are shown in tables \ref{tbl-performance-per-benchmark} and \ref{tbl-codesize-per-benchmark}.

\begin{figure}
 \centering
 \begin{minipage}{0.45\textwidth}
  \centering
  \makebox[\hsize][c]{\includegraphics[width=\myfiguresizeperformance]{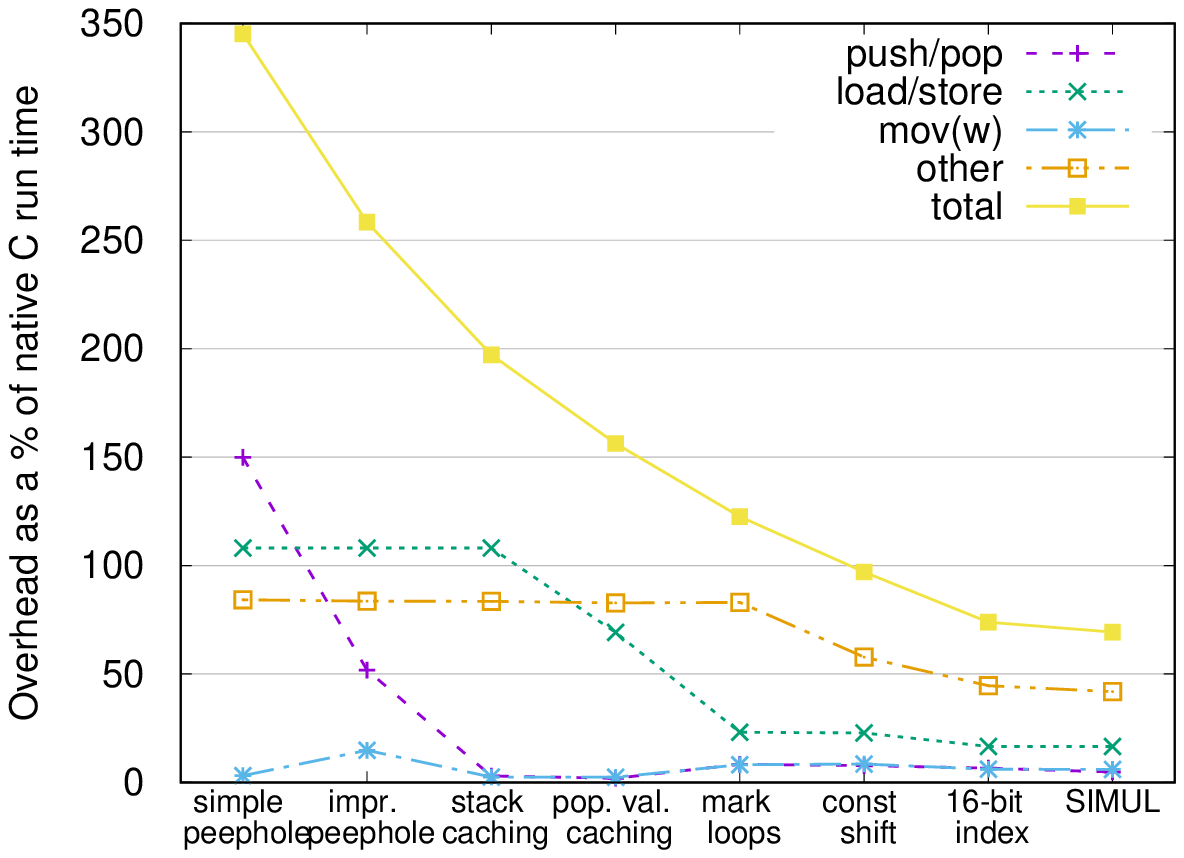}}
  \caption{Perf. overhead per category}
  \label{fig-performance-per-opcode-category}
 \end{minipage}\hfill
 \begin{minipage}{0.45\textwidth}
  \centering
  \makebox[\hsize][c]{\includegraphics[width=\myfiguresizeperformance]{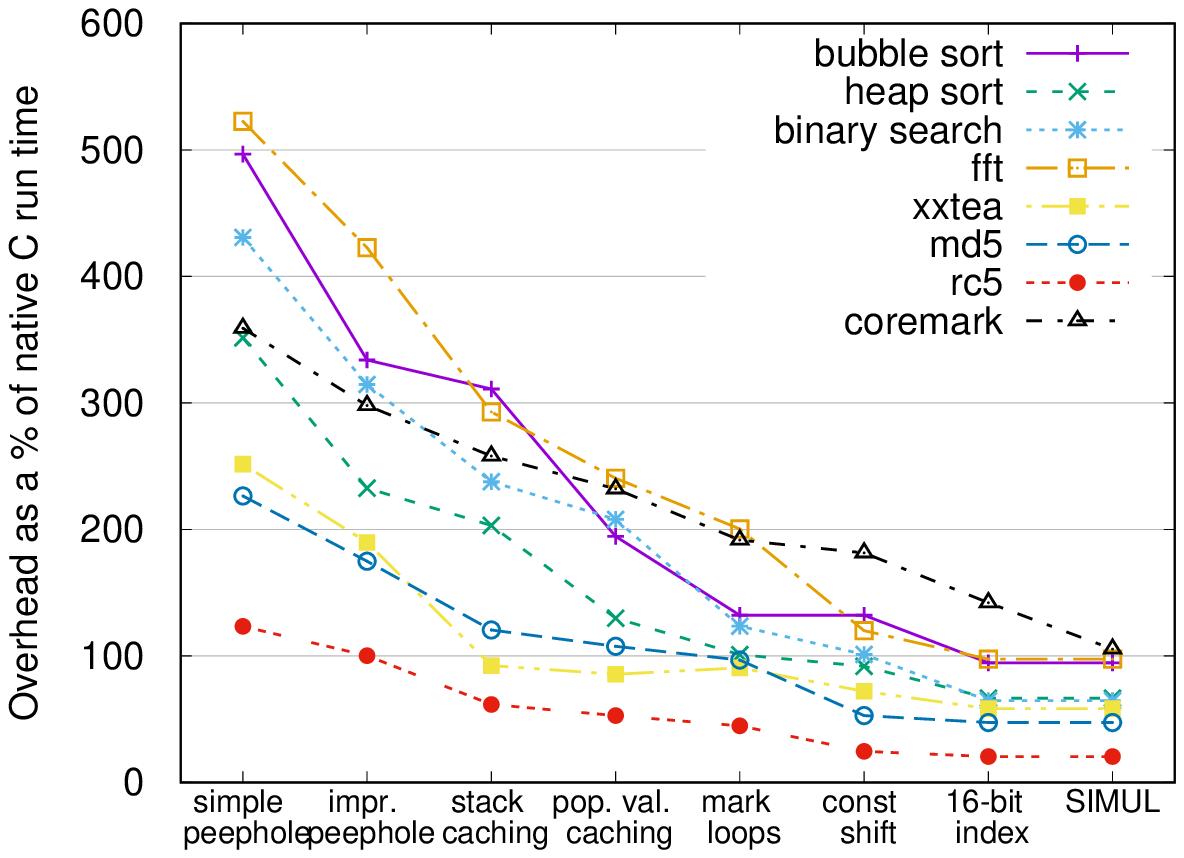}}
  \caption{Perf. overhead per benchmark}
  \label{fig-performance-per-benchmark}
 \end{minipage}
\end{figure}

In Figure \ref{fig-performance-per-opcode-category} we see how our optimisations combine to reduce performance overhead. We take the average of the 8 benchmarks, and show both the total overhead, and the overhead for each instruction category. Figure \ref{fig-performance-per-benchmark} shows the total overhead for each individual benchmark. We start with the original AOT approach with only the simple peephole optimiser, and then incrementally add each of our optimisations. The lightweight method call optimisation is already included in these results. Its effect will be examined in detail in Section \ref{sec-method-invocation}.

Using the simple optimiser, the types 1, 2 and 3 overhead are all significant, at 150\%,  108\%, and 84\% respectively. The basic approach does not have many reasons to emit a move, so we see that in some cases the AOT version actually spends fewer cycles on move instructions than the C version, resulting in small negative values. When we improve the peephole optimiser to include non-consecutive push/pop pairs, push/pop overhead drops by 98.1\% (of native C performance), but if the push and pop target different registers, they are replaced by a move instruction, and we see an increase of 11.7\% in move overhead. For a 16-bit value this takes 1 cycle (for a MOVW instruction), instead of 8 cycles for two pushes and two pops. The increase in moves shows most of the extra cases that are handled by the improved optimiser are replaced by a move instead of eliminated, since the 11.7\% extra move overhead matches a 93.6\% reduction in push/pop overhead.

Next we introduce stack caching to utilise all available registers and eliminate most of the push/pop instructions that cannot be handled by the improved optimiser. As a result the push/pop overhead drops to nearly 0, and so does the move overhead since most of the moves introduced by the peephole optimiser, are also unnecessary when using stack caching.

Having eliminated the type 1 overhead almost completely, we now add popped value caching to remove a large number of the unnecessary load instructions. This reduces the memory traffic significantly, as is clear from the reduced load/store overhead, while the other types remain stable. Adding the mark loops optimisation further reduces loads, and this time also stores, by pinning common variables to a register. But it uses slightly more move instructions, and the fact that we have fewer registers available for stack caching means we have to spill stack values to memory more often. While we save 46\% on loads and stores, the push/pop and move overhead both increase by 6\%.

Most of the push/pop and load/store overhead has now been eliminated and the type 3 overhead, unaffected by these optimisations, has become the most significant source of overhead. This type has many different causes, but we can eliminate half of it with our three instruction set optimisations. These optimisations, especially the 16-bit array index, also reduce register pressure, so we also see slight decreases in the other overhead types, although this is minimal in comparison. The CoreMark benchmark is the only one to do 16-bit to 32-bit multiplication, so the average performance improvement for \texttt{SIMUL} is small, but Table \ref{tbl-performance-per-benchmark} shows it is very significant for CoreMark.

Combined, these optimisations reduce performance overhead from 345\% to 69\% of native C performance.

\subsection{Code size}
\begin{table}[]
 \centering
 \caption{Code size data per benchmark}
 \label{tbl-codesize-per-benchmark}
 \small
 \scriptsize
 \setlength{\tabcolsep}{4pt}
 \makebox[\hsize][c]{\begin{tabular}{lrrrrrrrrrrrr}
\toprule
BENCHMARK                          & b.sort            &  h.sort           & b.srch            & fft               & xxtea             & md5               & rc5               & coremk            & \makebox[0.2mm]{}   & average           \\
\hline
\multicolumn{10}{l}{CODE SIZE (BYTES)} \\
\xxt JVM                           &                78 &               140 &                91 &               493 &               384 &              2986 &               457 &              5719 &     &                   \\
\xxt Native C                      &               150 &               416 &               212 &              1214 &              1442 &              9458 &               910 &             10388 &     &                   \\
\xxt AOT original                  &               520 &              1170 &               616 &              2694 &              3780 &             29362 &              4074 &             33668 &     &                   \\
\xxt AOT optimised                 &               344 &               738 &               450 &              1460 &              2268 &             14798 &              2140 &             25560 &     &                   \\
\hline
\multicolumn{10}{l}{CODE SIZE OVERHEAD BEFORE OPTIMISATIONS (\%)} \\
\xxt Total                         &             242.1 &             179.9 &             190.6 &             121.9 &             162.1 &             210.4 &             347.7 &             223.8 &     &             209.8 \\
  \xxxt push/pop                   & \xt          57.9 & \xt          61.2 & \xt          52.8 & \xt          55.7 & \xt         102.6 & \xt         133.1 & \xt         163.1 & \xt          74.5 &     & \xt          87.6 \\
  \xxxt load/store                 & \xt          89.5 & \xt          64.1 & \xt          69.8 & \xt          31.8 & \xt          28.4 & \xt          56.7 & \xt          67.9 & \xt          53.5 &     & \xt          57.7 \\
  \xxxt mov(w)                     & \xt           1.3 & \xt           1.4 & \xt           0.9 & \xt           0.3 & \xt           0.7 & \xt          -2.7 & \xt          -1.3 & \xt           1.4 &     & \xt           0.3 \\
  \xxxt other                      & \xt          93.4 & \xt          53.1 & \xt          67.0 & \xt          34.1 & \xt          30.4 & \xt          23.3 & \xt         118.0 & \xt          94.4 &     & \xt          64.2 \\
\multicolumn{10}{l}{CODE SIZE OVERHEAD REDUCTION PER OPTIMISATION (\%)} \\
\xxt Impr. peephole                &             -57.9 &             -41.1 &             -45.3 &             -26.5 &             -38.5 &             -54.3 &             -62.4 &             -30.7 &     &             -44.6 \\
\xxt Stack caching                 &             -13.1 &             -20.6 &             -24.5 &             -37.1 &             -56.1 &             -78.6 &            -106.4 &             -18.7 &     &             -44.4 \\
\xxt Pop. val. caching             &             -18.5 &             -27.8 &               0.0 &             -13.8 &              -6.2 &             -18.8 &             -17.8 &             -12.6 &     &             -14.4 \\
\xxt Mark loops                    &              -2.6 &              +4.8 &              +7.5 &              -5.9 &              +6.0 &              -1.1 &              -3.7 &              -3.8 &     &               0.2 \\
\xxt Const shift                   &               0.0 &              -2.4 &              -4.7 &              -5.3 &              +1.6 &              +4.0 &              -5.5 &              -1.1 &     &              -1.7 \\
\xxt 16-bit array index            &             -23.7 &             -16.2 &             -11.3 &             -13.0 &             -11.6 &              -5.1 &             -16.7 &              -8.8 &     &             -13.3 \\
\xxt SIMUL                         &               0.0 &               0.0 &               0.0 &               0.0 &               0.0 &               0.0 &               0.0 &              -2.3 &     &              -0.3 \\
\multicolumn{10}{l}{CODE SIZE OVERHEAD AFTER OPTIMISATIONS (\%)} \\
\xxt Total                         &             126.3 &              76.6 &             112.3 &              20.3 &              57.3 &              56.5 &             135.2 &             145.8 &     &              91.3 \\
  \xxxt push/pop                   & \xt          21.1 & \xt           5.7 & \xt           7.5 & \xt           0.0 & \xt          13.3 & \xt           0.0 & \xt           4.4 & \xt          19.4 &     & \xt           8.9 \\
  \xxxt load/store                 & \xt          31.6 & \xt          33.5 & \xt          47.2 & \xt           4.1 & \xt          14.8 & \xt          37.2 & \xt          25.3 & \xt          36.8 &     & \xt          28.8 \\
  \xxxt mov(w)                     & \xt           0.0 & \xt           3.8 & \xt           4.7 & \xt          -2.6 & \xt           2.5 & \xt          -1.8 & \xt          17.8 & \xt           5.4 &     & \xt           3.7 \\
  \xxxt other                      & \xt          73.7 & \xt          33.5 & \xt          52.8 & \xt          18.8 & \xt          26.6 & \xt          21.0 & \xt          87.7 & \xt          84.2 &     & \xt          49.8 \\
\bottomrule
\end{tabular}}

 \setlength{\tabcolsep}{6pt}
\end{table}

Next we examine the effects of our optimisations on code size. Two factors are important here: the size of the VM itself and the size of the code it generates.

\begin{figure}
 \centering
 \begin{minipage}{0.45\textwidth}
  \centering
  \makebox[\hsize][c]{\includegraphics[width=\myfiguresizecodesize]{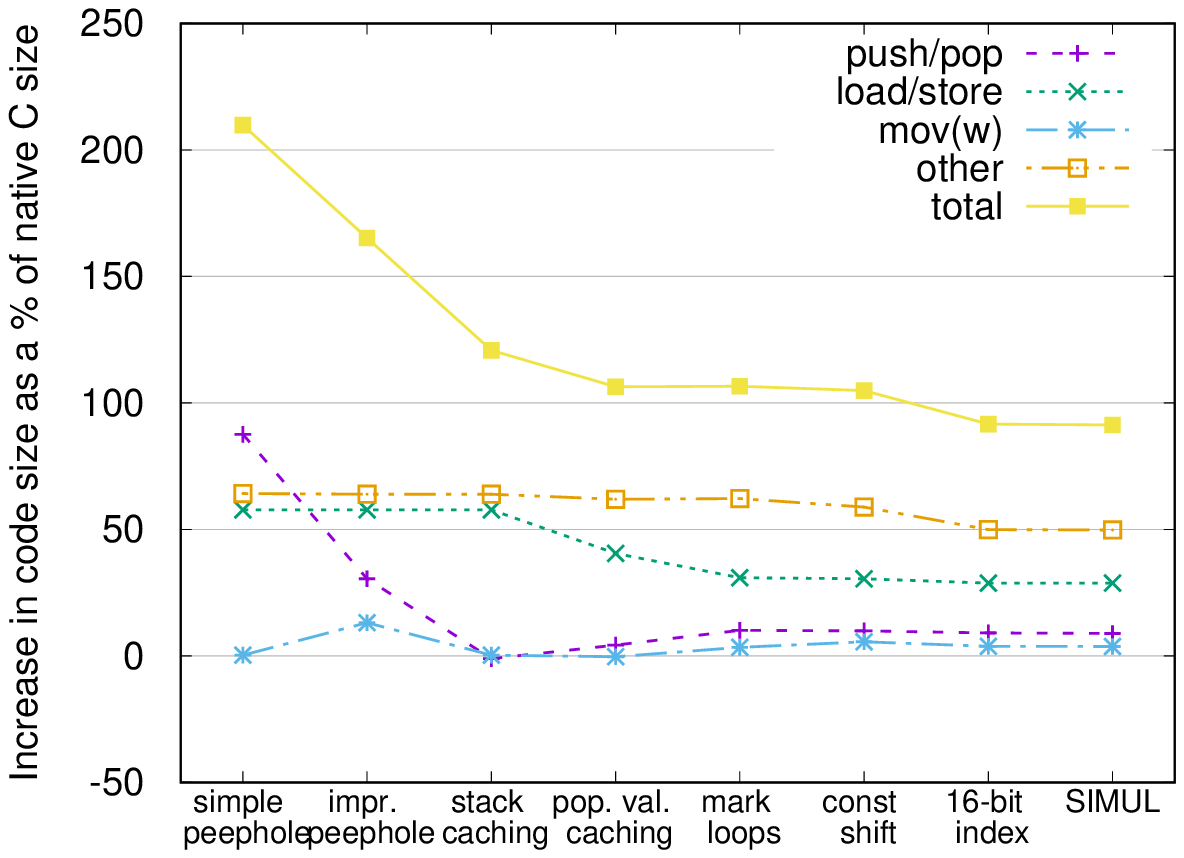}}
  \caption{Code size overhead per category}
  \label{fig-codesize-per-opcode-category}
 \end{minipage}\hfill
 \begin{minipage}{0.45\textwidth}
  \centering
  \makebox[\hsize][c]{\includegraphics[width=\myfiguresizecodesize]{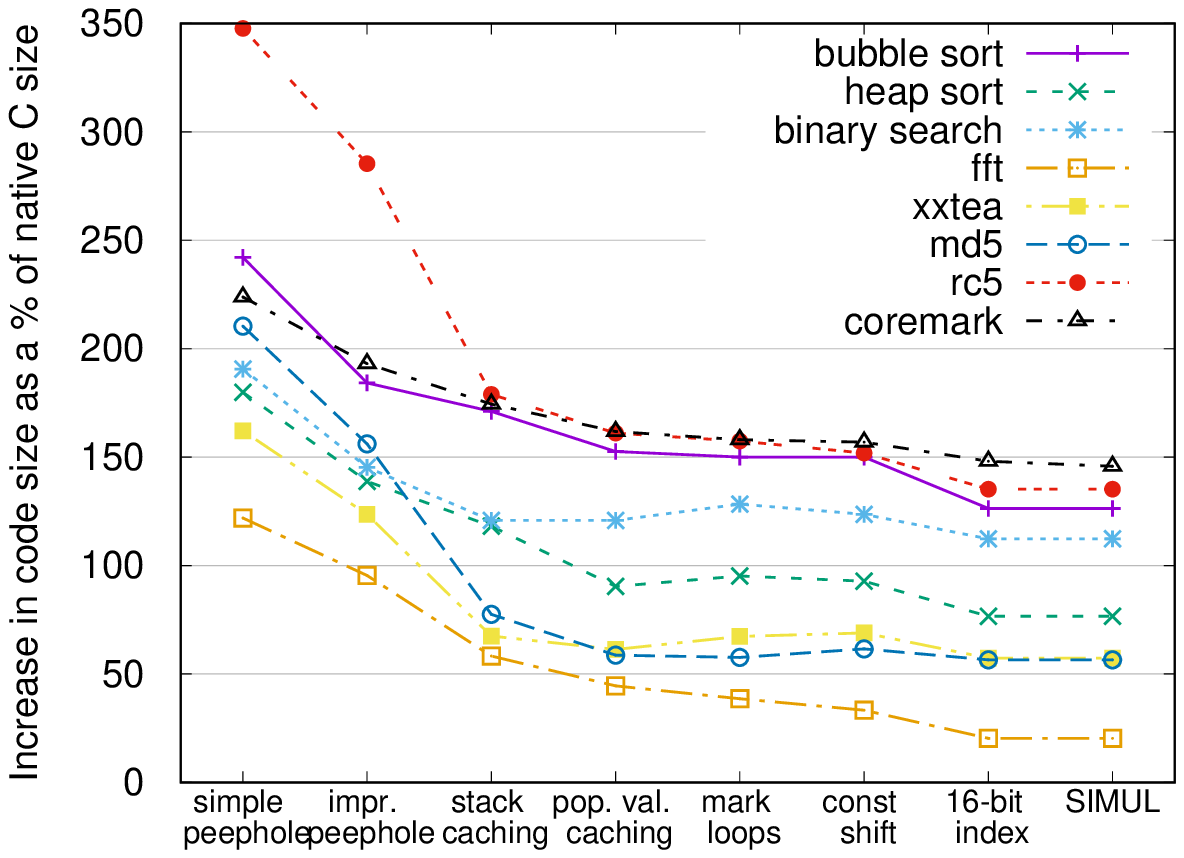}}
  \caption{Code size overhead per benchmark}
  \label{fig-codesize-per-benchmark}
 \end{minipage}
\end{figure}

The size overhead for the generated code is shown in figures \ref{fig-codesize-per-opcode-category} and \ref{fig-codesize-per-benchmark}, again split up per instruction category and benchmark respectively. For the first three optimisations, the two graphs follow a similar pattern as the performance graphs. These optimisations eliminate the need to emit certain instructions, which reduces code size and improves performance at the same time.

The mark loops optimisation moves loads and stores for pinned variables outside of the loop. This reduces performance overhead by 34\%, but the effect on code size varies per benchmark: some are slightly smaller, others slightly larger.

For each value that is live at the beginning of the loop, we need to emit the load before the mark loop block, so in terms of code size we only benefit if it is loaded more than once, and may actually lose some if it is then popped destructively, since we would need to emit a \texttt{mov}. Stores follow a similar argument. Also, for small methods the extra registers used may mean we have to save more call-saved registers in the method prologue. Finally, we get the performance advantage for each run-time iteration, but the effect on code size, whether positive or negative, only once.

The constant shift optimisation unrolls the loop that is normally generated for bit shifts. This significantly improves performance, but the effect on the code size depends on the number of bits to shift by. The constant load and loop take at least 5 instructions. In most cases the unrolled shifts will be smaller, but md5 actually shows a small 4\% increase in code size since it contains many shifts by a large number of bits.

Using 16-bit array indexes also reduces code size. The benchmarks here already have the manual code optimisations, so they use short index variables. This means the infuser will emit a \texttt{S2I} instruction to cast them to 32-bit ints if the array access instructions expect an int index. Not having to emit those when the array access instructions expect a 16-bit index, and the reduced work the access instruction needs to do, saves about 13\% code size overhead in addition to the 23\% reduction in performance overhead. Using 32-bit variables in the source code would also remove the need for \texttt{S2I} instructions, but the extra code to manipulate the index variable would make the net code size even larger.

\subsubsection{VM code size and break-even point}
These more complex code generation techniques do increase the size of our compiler. The first column in Table \ref{tbl-code-size-and-memory-consumption} shows the difference in code size between the AOT translator and Darjeeling's interpreter. The basic AOT approach is 6245B larger than the interpreter, and each of our optimisations adds a little to the size of the VM.

They also generate significantly smaller code. The second column shows the reduction in the generated code size compared to the baseline approach. Here we show the reduction in total size, as opposed to the overhead used elsewhere, to be able to calculate the break-even point. Using the improved peephole optimiser adds 278 bytes to the VM, but it reduces the size of the generated code by 14.4\%. If we have more than 1.9KB available to store user programmes, this reduction will outweigh the increase in VM size. Adding more complex optimisations further increases the VM size, but compared to the baseline approach, the break-even point is well within the range of memory typically available on a sensor node, peaking at at most 18.1KB.

As is often the case, there is a tradeoff between size and performance. The interpreter is smaller than each version of our AOT compiler, and Table \ref{tbl-codesize-per-benchmark} shows JVM bytecode is smaller than both native C and AOT compiled code, but the interpreter's performance penalty may be unacceptable in many cases. Using AOT compilation we can achieve adequate performance, but the most important drawback has been an increase in generated code size. These optimisations help to mitigate this drawback, and both improve performance, and allow us to load more code on a device.

For the smallest devices, or if we want to be able to load especially large programmes, we may decide to use only a selection of optimisations to limit the VM size and still get both a reasonable performance, and most of the code size reduction. Using only popped value stack caching reduces code size by 33.3\%, and results in a performance overhead of 156\%. The  16-bit array index optimisation should also be included, since this reduces the size of both the VM and the generated code.

\begin{table*}[]
\centering
\caption{Code size and memory consumption}
\label{tbl-code-size-and-memory-consumption}
\small
\makebox[\hsize][c]{\begin{tabular}{lrrrrrr}
\toprule
                      & size vs     & size vs  &            & AOT code  & break   & memory \\
                      & interpreter & baseline &            & reduction & even    & usage  \\
\hline
Baseline              &     6245 B  &          &            &           &         & 30 B   \\
Improved peephole     &     6523 B  &    278 B &  \scriptsize (+278)  &  -14.4\%  &  1.9 KB & 30 B   \\
Simple stack caching  &     7243 B  &    998 B &  \scriptsize (+720)  &  -28.6\%  &  3.5 KB & 41 B   \\
Popped value caching  &     8607 B  &   2362 B & \scriptsize (+1364)  &  -33.3\%  &  7.1 KB & 89 B   \\
Markloop              &    11903 B  &   5658 B & \scriptsize (+3296)  &  -33.2\%  & 17.0 KB & 98 B   \\
Const shift           &    12373 B  &   6128 B &  \scriptsize (+470)  &  -33.8\%  & 18.1 KB & 98 B   \\
16-bit array index    &    12353 B  &   6108 B &   \scriptsize (-20)  &  -38.0\%  & 16.1 KB & 98 B   \\
SIMUL                 &    12419 B  &   6174 B &   \scriptsize (+66)  &  -38.1\%  & 16.2 KB & 98 B   \\
Lightweight methods   &    12961 B  &   6716 B &  \scriptsize (+542)  &  -38.4\%  & 17.5 KB & 98 B   \\
\bottomrule
\end{tabular}}
\end{table*}

\subsubsection{VM memory consumption} The last column in Table \ref{tbl-code-size-and-memory-consumption} shows the size of the main data structure that needs to be kept in memory while translating a method. For the baseline approach we only use 30 bytes for a number of commonly used values such as a pointer to the next instruction to be compiled, the number of instructions in the method, etc. The simple stack caching approach adds a 11 byte array to store the state of each register pair we use for stack caching. Popped value caching adds two more arrays of 16-bit elements to store the value tag and age of each value. Mark loops only needs an extra 16-bit word to mark which registers are pinned, and a few other variables. Finally, the instruction set optimisations do not require any additional memory. In total, our compiler requires 98 bytes of memory during the compilation process.

\subsection{Benchmark details}
Next, we have a closer look at some of the benchmarks and see how the effectiveness of each optimisation depends on the characteristics of the source code. The first section of Table \ref{tbl-performance-per-benchmark} shows the distribution of the JVM instructions executed in each benchmark, and both the maximum and average number of bytes on the JVM stack. We can see some important differences between the benchmarks. While the benchmarks on the left are almost completely load/store bounded, towards the right the benchmarks become more computation intensive, spending fewer instructions on loads and stores, and more on math or bitwise operations. The left benchmarks have only a few bytes on the stack, but as the benchmarks contain more complex expressions, the number of values on the stack increases.

The second part of tables \ref{tbl-performance-per-benchmark} and \ref{tbl-codesize-per-benchmark} first shows the overhead before optimisation, split up in the four instruction categories. We then list the effect of each optimisation on the total overhead. Finally we show the overhead per category after applying all optimisations.

The improved peephole optimiser and stack caching both target the push/pop overhead. Stack caching can eliminate almost all, and replaces the need for a peephole optimiser, but it is interesting to compare the two. The improved peephole optimiser does well for the simple benchmarks like sort and search, leaving less overhead to remove for stack caching. Moving to the right, the more complicated expressions mean there is more distance between a push and a pop, leaving more cases that cannot be handled by the peephole optimiser, and replacing it with stack caching yields a big improvement.

The benchmarks on the left spend more time on load/store instructions. This results in higher load/store overhead, and the two optimisations that target this overhead, popped value caching and mark loops, have a big impact. For the computation intensive benchmarks on the right, the load/store overhead is much smaller, but the higher stack size means stack caching is very important for these benchmarks.

The first seven benchmarks are smaller benchmarks that can highlight certain specific aspects of our approach, the CoreMark benchmark represents larger sensor node applications, and is a mix of different types of processing. As a result, it is an average case in almost every row in Table \ref{tbl-performance-per-benchmark}. The reason it ends up being the slowest after all optimisations was discussed in \ref{sec-evaluation-coremark-unfair-optimisations}. With the 'unfair' optimisations described there, CoreMark's performance overhead would be 61\%, very close to the average of the other benchmarks.

\textbf{Bit shifts} Interestingly, the reason fft is the slowest, is similar to the reason rc5 is fastest: they both spend a large amount of time doing bit shifts. Rc5 shifts by a variable, but large number of bits. Only 8.0\% of the executed JVM instructions are bit shifts, but they account for 71\% of the execution time in the optimised version. For these variable bit shifts, our translator and \texttt{avr-gcc} generate a similar loop, so the two share a large constant factor.

On the other hand fft is a hard case because it does many constant shifts by exactly 6 bits. For these, our VM simply emits 6 single shifts, which is slower than the special case \texttt{avr-gcc} emits for shifts by exactly 6 bits.  While we could do the same, we feel this special case is too specific to include in our VM.

\textbf{Bubble sort} Next we look at bubble sort in some more detail. After optimisation, we see most of the stack related overhead has been eliminated and of the 94.6\% remaining performance overhead, most is due to other sources. For bubble sort there is a single, clearly identifiable source. When we examine the detailed trace output, this overhead is largely due to \texttt{ADD} instructions, but bubble sort hardly does any additions. This is a good example of how the simple JVM instruction set leads to less efficient code. To access an array we need to calculate the address of the indexed value, which takes one move and seven additions for an array of ints. This calculation is repeated for each access, while the C version has a much more efficient approach, using the auto-increment version of the AVR's LD and ST instructions to slide a pointer over the array. Of the remaining 94.6\% overhead, 73\% is caused by these address calculations.

\begin{figure}
 \centering
 \begin{minipage}{0.45\textwidth}
  \centering
  \makebox[\hsize][c]{\includegraphics[width=\myfiguresizexxtea]{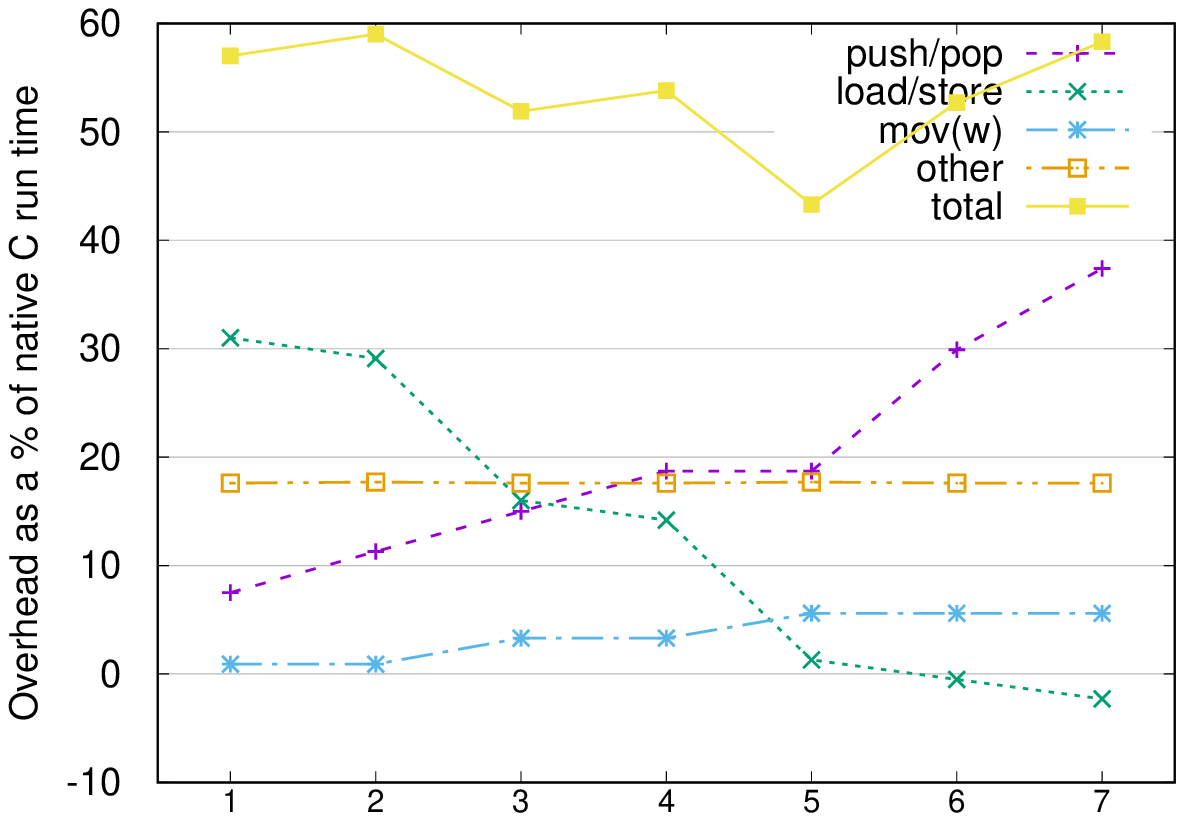}}
  \caption{Xxtea performance overhead for different number of pinned register pairs}
  \label{fig-performance-pinnedregs-xxtea-per-opcode-category}
 \end{minipage}\hfill
 \begin{minipage}{0.45\textwidth}
  \centering
  \makebox[\hsize][c]{\includegraphics[width=\myfiguresizexxtea]{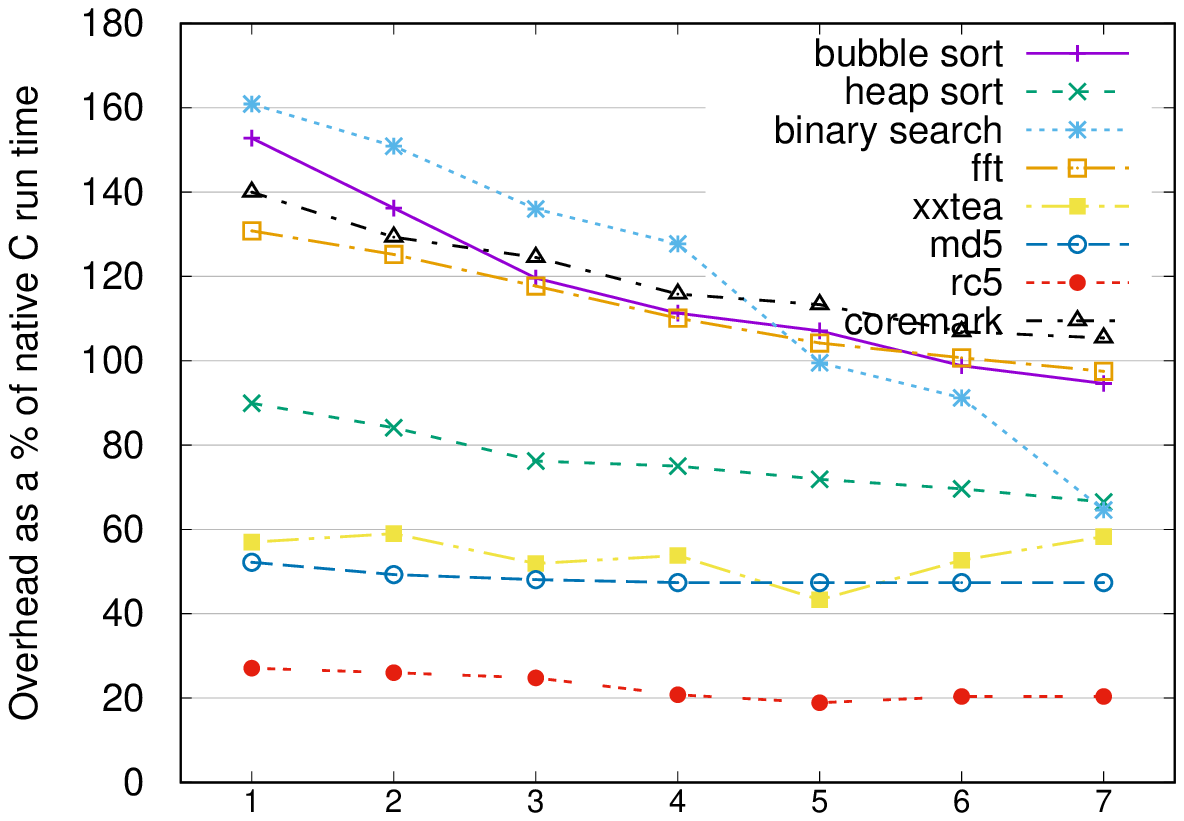}}
  \caption{Per benchmark performance overhead different number of pinned register pairs}
  \label{fig-performance-pinnedregs-per-benchmark}
 \end{minipage}
\end{figure}

\textbf{Xxtea and the mark loops optimisation} Perhaps the most interesting benchmark is xxtea. Its high average stack depth means popped value caching does not have much effect: most registers are used for real stack values, leaving few chances to reuse a value that was previously popped from the stack. 

When we apply the mark loops optimisation, performance actually degrades by 5.1\%, and code size overhead increases 6\%! Here we have an interesting tradeoff: if we use a register to pin a variable, accessing that variable will be cheaper, but this register will no longer be available for stack caching, so more stack values may have to be spilled to memory.

For most benchmarks the maximum of 7 register pairs to pin variables to was also the best option. At a lower average stack depth, the fewer number of registers available for stack caching is easily compensated for by the cheaper variable access. For xxtea however, the cost of spilling more stack values to memory outweighs the gains of pinning more variables when too many variables are pinned. Figure \ref{fig-performance-pinnedregs-xxtea-per-opcode-category} shows the overhead for xxtea from the different instruction categories. When we increase the number of register pairs used to pin variables from 1 to 7, the load/store overhead steadily decreases, but the push/pop and move overhead increase. The optimum is at 5 pinned register pairs, at which the total overhead is only 43\%, instead of 58\% at 7 pinned register pairs.

 Interestingly, when we pin 7 pairs, the AOT version actually does fewer loads and stores than the C compiler. Under high register pressure the C version may spill a register value to memory and later load it again, adding extra load/store instructions. When the AOT version pins too many registers, it will also need to spill values, but this adds push/pop instructions instead of loads/stores.

Figure \ref{fig-performance-pinnedregs-per-benchmark} shows the performance for each benchmark, as the number of pinned register pairs is increased. The three benchmarks that stay stable or even slow down when the number pinned pairs is increased beyond 5 are exactly the benchmarks that have a high stack depth: xxtea, md5 and rc5. It should be possible to develop a simple heuristic to allow the VM to make a better decision on the number of registers to pin. Since our current VM always pins 7 pairs, we used this as our end result and leave this heuristic to future work.

\subsection{Method invocation}
\label{sec-method-invocation}
\begin{table*}[]
\centering
\caption{Methods per benchmark and relative performance for normal, lightweight invocation, and inlining. Highlights indicate changes from the versions used to obtain the results in the previous sections.}
\label{tbl-benchmark-methods}
\scriptsize
\makebox[\hsize][c]{\begin{tabular}{lllllll}
\toprule
                             & \# calls                     & C                 & Java                          & Java                            & Java                            \\
                             &                              &                   & Base version                  & Alternative version             & Using normal method calls       \\
\hline
\\
CoreMark \\
ee\_isdigit                  & \multicolumn{1}{r}{3920}     & normal (inlined)  & manually inlined              & \tblhighlight lightweight (JVM) & manually inlined                \\
core\_state\_transition      & \multicolumn{1}{r}{1024}     & normal            & lightweight                   & lightweight                     & \tblhighlight normal            \\
crcu8                        & \multicolumn{1}{r}{584}      & normal (inlined)  & lightweight                   & lightweight                     & \tblhighlight normal            \\
crcu16                       & \multicolumn{1}{r}{292}      & normal            & lightweight                   & lightweight                     & \tblhighlight normal            \\
calc\_func                   & \multicolumn{1}{r}{220}      & normal            & lightweight                   & lightweight                     & \tblhighlight normal            \\
compare\_idx                 & \multicolumn{1}{r}{209}      & normal (inlined)  & normal (virtual)              & normal (virtual)                & normal (virtual)                \\
core\_list\_find             & \multicolumn{1}{r}{206}      & normal            & lightweight                   & lightweight                     & \tblhighlight normal            \\
compare\_complex             & \multicolumn{1}{r}{110}      & normal            & normal (virtual)              & normal (virtual)                & normal (virtual)                \\
crcu32                       & \multicolumn{1}{r}{64}       & normal            & lightweight                   & lightweight                     & \tblhighlight normal            \\
matrix\_sum                  & \multicolumn{1}{r}{16}       & normal            & lightweight                   & lightweight                     & \tblhighlight normal            \\
others (<16 calls each)      & \multicolumn{1}{r}{39}       & normal            & normal                        & normal                          & normal                          \\
\\
\emph{cycles}                &                              &                   & \multicolumn{1}{r}{3482185}   & \multicolumn{1}{r}{3639967}     & \multicolumn{1}{r}{5030231}     \\
\emph{overhead v native C}   &                              &                   & \multicolumn{1}{r}{105.4\%}   & \multicolumn{1}{r}{114.7\%}     & \multicolumn{1}{r}{196.7\%}     \\
\emph{code size}             &                              &                   & \multicolumn{1}{r}{25560}     & \multicolumn{1}{r}{25576}       & \multicolumn{1}{r}{26282}       \\
\\
\hline
\\
FFT \\
FIX\_MPY                     & \multicolumn{1}{r}{768}      & marked inline     & manually inlined              & \tblhighlight lightweight (JVM) & \tblhighlight normal            \\
SIN8                         & \multicolumn{1}{r}{63}       & marked inline     & manually inlined              & \tblhighlight ProGuard inlined  & \tblhighlight ProGuard inlined  \\
COS8                         & \multicolumn{1}{r}{63}       & marked inline     & manually inlined              & \tblhighlight ProGuard inlined  & \tblhighlight ProGuard inlined  \\
\\
\emph{cycles}                &                              &                   & \multicolumn{1}{r}{78241}     & \multicolumn{1}{r}{113611}      & \multicolumn{1}{r}{562650}      \\
\emph{overhead v native C}   &                              &                   & \multicolumn{1}{r}{97.5\%}    & \multicolumn{1}{r}{186.8\%}     & \multicolumn{1}{r}{1320.4\%}    \\
\emph{code size}             &                              &                   & \multicolumn{1}{r}{1460}      & \multicolumn{1}{r}{1410}        & \multicolumn{1}{r}{1530}        \\
\\
\hline
\\
heap sort \\
SWAP                         & \multicolumn{1}{r}{1642}     & \#define          & manually inlined              & manually inlined                & manually inlined                \\
siftDown                     & \multicolumn{1}{r}{383}      & normal            & lightweight                   & \tblhighlight manually inlined  & \tblhighlight normal            \\
\\
\emph{cycles}                &                              &                   & \multicolumn{1}{r}{289845}    & \multicolumn{1}{r}{286749}      & \multicolumn{1}{r}{563967}      \\
\emph{overhead v native C}   &                              &                   & \multicolumn{1}{r}{66.5\%}    & \multicolumn{1}{r}{64.7\%}      & \multicolumn{1}{r}{223.9\%}     \\
\emph{code size}             &                              &                   & \multicolumn{1}{r}{738}       & \multicolumn{1}{r}{926}         & \multicolumn{1}{r}{758}         \\
\bottomrule
\end{tabular}}
\end{table*}

Most of our benchmarks consist of only a single method. The three small functions in the FFT benchmark were inlined by the C compiler, so we manually inlined them in the Java version. Heap sort does contain a real method call: it consist of a main loop, repeatedly calling the \texttt{siftDown} method. CoreMark is a much more extensive benchmark consisting of many methods.

In this section we will examine the effect of the lightweight method calls on these three benchmarks, compared to inlined code and normal method calls.

In Table \ref{tbl-benchmark-methods} we see the most frequently called methods of the CoreMark, FFT and heap sort benchmarks, and the number of times they are called in a single run. Next, we list the way they are implemented in C. CoreMark only defines normal functions, which are inlined by \texttt{avr-gcc} in three cases. FFT contains 3 methods marked with the \texttt{inline} compiler hint, which was followed by \texttt{avr-gcc}. Finally heap sort uses just one extra function, and a macro to swap two array elements.

The Java base version column shows the way these functions are implemented in the Java versions of our benchmarks. We manually inlined C macros, and the smallest functions that were inlined by the C compiler. The most commonly called methods were transformed to lightweight methods, simply by adding the \texttt{@Lightweight} annotation where possible. For Java versions of the \texttt{compare\_idx} and \texttt{compare\_complex} methods, this was not possible since we do not support lightweight virtual methods.

In the next two columns we vary these choices slightly to examine the effect of our lightweight methods.

For the CoreMark benchmark, we first replace the inlined implementation of the most frequently called method with a lightweight version. \texttt{ee\_isdigit} returns true if a \texttt{char} passed to it is between \texttt{'0'} and \texttt{'9'}. Since this is a very trivial method, we manually coded the lightweight method to use only the stack and no local variables. This slowed the benchmark down by 4.5\%, adding 157,782 cycles. Since the method is called 3920 times, this corresponds to an overhead of about 40 cycles, which is on the high side for such a small method.

Here we see another overhead from using a lightweight method that's hard to quantify: the boolean result of \texttt{ee\_isdigit} is used to decide an \texttt{if} statement. When we inline the code, the VM can directly branch on the result of the expression \texttt{(c>='0' \&\& c<='9')}, but the lightweight method first has to return a boolean, which is then tested again after the lightweight call returns.

Next, we see what the performance would be without lightweight methods, and all methods, except the manually inlined \texttt{ee\_isdigit}, have to be implemented as normal Java methods. This adds a total of 1,548,046 cycles, making it almost 1.5 times slower than the lightweight methods version. Spread over 2406 calls, this means the average method invocation added over 643 cycles, which is within the range predicted in Section \ref{sec-optimisations-method-calls}.

The FFT benchmark has a much lower running time than CoreMark, but still does 894 function calls. In the C and normal Java versions these are inlined. When we change them all to normal Java methods, ProGuard will automatically inline the \texttt{SIN8} and \texttt{COS8} methods, adding only a minimal overhead, but the \texttt{FIX\_MPY} method is too large for ProGuard to inline. If we mark it \texttt{@Lightweight} the large number of calls relative to the total running time means the average overhead of over 40 cycles per invocation slows down the benchmark by 45\%. Without lightweight methods, this would be as high as 619\%

Finally, for the heap sort benchmark we normally use a lightweight method for \texttt{siftDown}. In the second version we see that, like in the CoreMark example, the difference between inlining and the lightweight method is small: we only gain 1\% by manual inlining. However, the benchmark runs almost twice as long when we use a normal method call instead of a lightweight method. A significant increase, but less than FFT since heap sort does half as many calls and has a higher total running time to spread the call overhead.

In terms of code size, we can see normal methods take slightly more space than a lightweight method. Listing \ref{lst-comparison-lightweight-and-normal-invocation} showed that the invocation is more complex for normal methods, and in addition the method prologue and epilogue are longer.

The difference between inlining and lightweight methods is less clear. For the smallest of methods, such as CoreMark's \texttt{ee\_isdigit}, the inlined code is smaller than the call, but the heap sort benchmark shows that inlining larger methods can result in significantly larger code.

As these three examples show, using lightweight methods gives us an option in-between a normal method call and inlining. This avoids most of the overhead of a normal method call, and the potential size increase of inlining.

\section{Conclusions and future work}
A major problem for sensor node VMs has been performance. Most interpreters are between one to two orders of magnitude slower than native code, leading to both lower maximum throughput and increased energy consumption.

Previous work on AOT translation to native code by Ellul and Martinez \cite{Ellul:2010iw} improves performance, but still a significant overhead remains, and the tradeoff is that the resulting native code takes up much more space, limiting the size of programmes that can be loaded onto a device. For the CoreMark benchmark, the performance is 9x slower than native C, and the code 3.5 times larger.

In this paper, we presented the complete set of techniques we developed to mitigate this code size overhead and to further improve performance. We evaluated their effectiveness using a set of benchmarks, some with specific characteristics to highlight the results in more extreme conditions, and include the larger CoreMark benchmark to represent the average behaviour of larger sensor node applications. Combined, our optimisations result in a compiler that produces code that is on average only 1.7 times slower and 1.9 times larger than optimised C.

These optimisations do increase the size of our VM, but the break-even point at which this is compensated for by the smaller code it generates, is well within the range of programme memory typically available on a sensor node. This leads us to believe that these optimisations will be useful in many scenarios, and make using a VM a viable option for a wider range of applications.

Many opportunities for future work remain. In this paper we focus on techniques for the sensor node side, but a future VM should come with a better optimising infuser on the host to prepare better quality bytecode. This infuser should also support inlining small methods as efficiently as manual inlining, and in most cases automatically determine which methods should be made lightweight.

For the mark loops optimisation, a heuristic is needed to make a better decision on the number of registers to pin, and we can consider applying this optimisation to other blocks that have a single point of entry and exit as well. Since supporting preemptive threads is expensive to implement without the interpreter loop as a place to switch threads, we believe a cooperative concurrency model where threads explicitly yield control is more suitable for sensor nodes using AOT, and we are working on building this on top of Darjeeling's existing thread support.

A more general question is what the most suitable architecture and instruction set is for a VM on tiny devices. Hsieh et al. note that the performance problem lies in the mismatch between the VM and the native machine architecture \cite{Hsieh:1996cy}. In this paper we presented a number of modifications to the bytecode format to make it better suited for use on a sensor now, but ultimately we believe JVM is not the best choice for a sensor node VM. It has some advanced features, such as exceptions, preemptive threads, and garbage collection, which add complexity but may not be necessary on a tiny device. At the same time, there is no support for constant data, which is common in embedded code: a table with sine wave values in the fft benchmark is represented as a normal array at run-time, using up valuable memory.
We may also consider extending the bytecode with instructions to express common operations more efficiently. For example, an instruction to loop over an array such as the one found in Lua \cite{Lua:2005} would allow us to generate more efficient code and eliminate most of the remaining overhead in the bubble sort benchmark.

Our reason to use JVM is the availability of a lot of infrastructure to build on. Like Hsieh et al., we do not claim that Java is the best answer for a sensor node VM, but we believe the techniques presented here will be useful in developing better sensor node VMs, regardless of the exact instruction set used.

One important question that should be considered is whether that instruction set should be stack-based or register-based. Many modern bytecode formats are register-based, and a number of publications report on the advantages of this approach \cite{Zhang:2012wf, Shi:2005ba}. However, these tradeoffs are quite different for a powerful JIT compiler, and a resource-constrained VM. When working with tiny devices, an important advantage of a stack-based architecture is its simplicity, and our results here show that much of the overhead associated with the stack-based approach can be eliminated during the translation process.

\section{Acknowledgements}
This research was supported in part by the Ministry of Science and Technology of Taiwan (MOST 105-2633-E-002-001), National Taiwan University (NTU-105R104045), Intel Corporation, and Delta Electronics.

\balance
\bibliographystyle{abbrv}
\bibliography{references}
    
\end{document}